\documentclass[journal=jacsat,manuscript=article]{achemso}
\usepackage[version=3]{mhchem}
\usepackage{hyperref}
\usepackage{amsmath}
\usepackage{xcolor}
\usepackage{booktabs}
\usepackage{xr}
\usepackage{siunitx}
\usepackage[inkscapelatex=false]{svg}

\newcolumntype{H}{>{\setbox0=\hbox\bgroup}c<{\egroup}@{}}

\newif\ifhighlightnumbers
\highlightnumbersfalse
\newcommand{\numericalresult}[1]{\ifhighlightnumbers \textcolor{red}{#1}\else #1\fi}

\newif\ifhighlightchanges
\highlightchangesfalse
\newcommand{\changed}[1]{\ifhighlightchanges \textcolor{red}{#1}\else #1\fi}

\newif\ifhighlightnewchanges
\highlightnewchangesfalse
\newcommand{\newchanges}[1]{\ifhighlightnewchanges \textcolor{red}{#1}\else #1\fi}

\newif\ifhighlightthirdchanges
\highlightthirdchangesfalse
\newcommand{\thirdchanges}[1]{\ifhighlightthirdchanges \textcolor{red}{#1}\else #1\fi}

\newif\ifhighlightfourthchanges
\highlightfourthchangesfalse
\newcommand{\fourthchanges}[1]{\ifhighlightfourthchanges \textcolor{red}{#1}\else #1\fi}

\makeatletter
\newcommand*{\addFileDependency}[1]{
  \typeout{(#1)}
  \@addtofilelist{#1}
  \IfFileExists{#1}{}{\typeout{No file #1.}}
}
\makeatother

\newcommand*{\myexternaldocument}[1]{%
    \externaldocument{#1}%
    \addFileDependency{#1.tex}%
    \addFileDependency{#1.aux}%
}

\myexternaldocument{nmr-supplement}

\author{Frank Hu}
\affiliation
{Department of Chemistry, Stanford University, Stanford, California 94305, United States}
\author{Jonathan M. Tubb}
\affiliation
{Department of Chemistry, Stanford University, Stanford, California 94305, United States}
\author{Dimitris Argyropoulos}
\affiliation
{ACD/Labs, Toronto, Ontario M5C 1B5, Canada}
\author{Sergey Golotvin}
\affiliation
{ACD/Labs, Toronto, Ontario M5C 1B5, Canada}
\author{Mikhail Elyashberg}
\affiliation
{ACD/Labs, Toronto, Ontario M5C 1B5, Canada}
\author{Grant M. Rotskoff}
\email{rotskoff@stanford.edu}
\affiliation
{Department of Chemistry, Stanford University, Stanford, California 94305, United States}
\author{Matthew W. Kanan}
\email{mkanan@stanford.edu}
\affiliation
{Department of Chemistry, Stanford University, Stanford, California 94305, United States}
\author{Thomas E. Markland}
\email{tmarkland@stanford.edu}
\affiliation
{Department of Chemistry, Stanford University, Stanford, California 94305, United States}

\title[NMR structure elucidation]
  {Pushing the limits of one-dimensional NMR spectroscopy for automated structure elucidation using artificial intelligence}

\abbreviations{IR,NMR}
\keywords{American Chemical Society, \LaTeX}

\begin{document}

\begin{abstract}
One-dimensional NMR spectroscopy is one of the most widely used techniques for the characterization of organic compounds and natural products. For molecules with up to 36 non-hydrogen atoms, the number of possible structures has been estimated to range from $10^{20} - 10^{60}$. The task of determining the structure (formula and connectivity) of a molecule of this size using only its one-dimensional \textsuperscript{1}H and/or \textsuperscript{13}C NMR spectrum, i.e. \textit{de novo} structure generation, thus appears completely intractable. Here we show how it is possible to achieve this task for systems with up to 40 non-hydrogen atoms across the full elemental coverage typically encountered in organic chemistry (C, N, O, H, P, S, Si, B, and the halogens) using a deep learning framework, thus covering a vast portion of the drug-like chemical space. Leveraging insights from natural language processing, we show that our transformer-based architecture predicts the correct molecule with \numericalresult{60.4\%} accuracy within the first 15 predictions using only the \textsuperscript{1}H and \textsuperscript{13}C NMR spectra, thus overcoming the combinatorial growth of the chemical space while also being extensible to experimental data via fine-tuning.
\end{abstract}

\section{Introduction}

One-dimensional nuclear magnetic resonance spectroscopy (1D NMR) is a workhorse method for structure elucidation in molecular sciences due to the relative ease of acquiring a spectrum and the richness of its information content. However, elucidation of the molecular structure (formula and connectivity) from 1D NMR data alone without information such as the molecular formula—i.e., \textit{de novo} structure generation—is intractable for all but the smallest molecules owing to the combinatorial scaling of chemical space with the number of atoms. For larger molecules, 1D NMR spectra must be used in combination with other information or more advanced characterization techniques such as 2D NMR (\newchanges{Heteronuclear Multiple Bond Correlation (HMBC), Correlated Spectroscopy (COSY), etc.) and High Resolution Mass Spectrometry (HR-MS)}, the analysis of which can be automated by the computer-assisted structure elucidation (CASE) approach\cite{elyashberg_computer_2021, elyashberg_enhancing_2023}. For example, for molecules with up to $\sim$36 heavy (non-hydrogen) atoms, a limit that encompasses the majority of drug-like molecules, $10^{20}$\cite{ertl_cheminformatics_2003} to $10^{60}$\cite{bohacek_art_1996,polishchuk_estimation_2013} possible molecules have been estimated to exist. An efficient automated framework that overcomes this formidable scaling challenge by correctly predicting the molecular structure from the vast chemical space of possible structures using only 1D NMR spectra would remove a bottleneck in the chemical discovery process and accelerate the synthesis and characterization of novel compounds. 

With machine learning (ML) playing an increasingly central role in scientific discovery, much effort has been devoted towards building ML-based tools and workflows for accelerating and automating the analysis of spectroscopic data. 
This includes ML-based methods to: perform the forward prediction of NMR chemical shifts\cite{binev_prediction_2007, meiler_proshift_nodate,gao_general_2020,li_highly_2024,han_scalable_2022,kuhn_building_2008,williamson_chemical_2024,guan_real-time_2021,gerrard_impression_2020,paruzzo_chemical_2018} (i.e. the structure-to-spectrum task) which can be used to aid spectrum matching approaches for structure elucidation\cite{lemm_impact_2024,howarth_dp4-ai_2020,priessner_hsqc_2024}; detect chemical fragments (substructures) from spectra\cite{li_identifying_2022,specht_automated_2024,huang_framework_2021,zhao_machine_2025,fine_spectral_2020,liu_automated_2025}; correlate different spectral features and regions with molecular motifs\cite{schilter_unveiling_2023}; and iteratively construct molecules through search algorithms\cite{huang_framework_2021} or ML-guided generative processes\cite{sridharan_deep_2022,devata_deepspinn_2024}. However, these approaches stop short of the most ambitious, but most challenging, task of elucidating the full molecular structure using only 1D NMR spectra as input (i.e. the spectrum-to-structure task): spectrum matching approaches require candidate structures to query against\cite{lemm_impact_2024,howarth_dp4-ai_2020,priessner_hsqc_2024}, substructure elucidation methods stop short of elucidating full molecular structures\cite{li_identifying_2022,specht_automated_2024,huang_framework_2021,zhao_machine_2025,fine_spectral_2020,liu_automated_2025}, and iterative molecular search and construction methods\cite{huang_framework_2021,sridharan_deep_2022,devata_deepspinn_2024} become infeasible when applied to large molecules since the chemical space grows combinatorially with the number of atoms. 

To achieve automated end-to-end structure elucidation, i.e., direct prediction of structure from spectra without intermediate steps, recent approaches have leveraged developments in deep learning by reframing the structure elucidation as a graph learning problem based on shielding constants\cite{sapegin_structure_2024} or an image recognition problem using 2D NMR and convolutional neural networks\cite{kim_deepsat_2023}. Transformer architectures\cite{vaswani_attention_2017} and their natural synergy with textual representations of molecules through SMILES\cite{weininger_smiles_1988} strings have led to a surge in approaches that exploit the power of transformers to capture subtle long-range correlations across long context strings, e.g. connections between disperse chemical regions of a molecule. These approaches vary widely in terms of the spectroscopic data sources used and the overall workflow. For instance, it has been shown that integer sequence representations of IR spectra\cite{alberts_leveraging_2024} or textual representations of NMR spectra\cite{alberts_learning_2023} can be used in concert with conditioning information, such as the molecular formula or confirmed molecular fragments\cite{yao_conditional_2023,sun_cross-modal_2024}, to directly predict a molecule's SMILES string. However, such conditioning information is not always readily available because of instrument or resource constraints, and these approaches require a significant amount of preprocessing of the spectra to create text- or sequence-based representations, which is time-consuming and can introduce biases through the annotation process. Genetic algorithms\cite{mirza_elucidating_2024} and diffusion processes\cite{yang_diffnmr_2026}, where molecules are iteratively updated based on previous generations of molecules and a stochastic differential equation-based denoising process, respectively, have also used transformers as encoders for guiding their generative processes. However, these \changed{iterative} approaches leverage additional context, including the molecular formula, which significantly simplifies the structure elucidation process while also suffering from high prediction costs.

The explosion of generalist large language models (LLMs) has led to the investigation of whether the current generation of these ``off-the-shelf" models can achieve accurate results on structure elucidation and spectrum reasoning tasks~\cite{yang_spectrumworld_2025,priessner_enhancing_2025}. However, it has been demonstrated that reasoning capabilities are essential for both these tasks, and reasoning models incur high inference costs due to the many additional tokens that must be processed for each input. The promising performance of generalist transformer-based architectures for the structure elucidation task has inspired others to create more specialized frameworks that use transformers as multimodal structure elucidators that \changed{take} numerous different spectra such as 2D/1D NMR, MS, IR, and UV-vis\cite{priessner_enhancing_2024,tan_transformer_2025,xu_spectre_nodate,alberts_automated_2025} to generate predictions of the molecular structure. However, owing to their multimodal nature, they require multiple pieces of increasingly intricate spectroscopic data for each molecule. 

An end-to-end structure elucidation framework that only requires routine 1D NMR spectra would provide a more compute- and data-efficient approach to the structure elucidation problem that avoids the expense of acquiring more complicated spectroscopic data, the necessity of often unavailable context, and the prediction cost of using generalist LLMs. Here we develop a transformer-based ML framework for solving the most challenging version of the structure elucidation problem, converting \textsuperscript{1}H and/or \textsuperscript{13}C NMR spectra directly into the molecular structure with minimal preprocessing and without the use of the molecular formula or any additional context. We use a powerful pretraining technique that leverages Morgan fingerprints\cite{rogers_extended-connectivity_2010} to adapt our transformer model to molecular representation learning and show that we can successfully reconstruct a molecule from its Morgan fingerprint (substructure-to-structure) with \numericalresult{97.8\%} accuracy within the first 15 predictions when testing on molecules with up to 40 heavy atoms and with the full elemental coverage typically encountered in organic chemistry (C, N, O, H, P, S, Si, B, and halogens). We integrate our pretrained transformer into an end-to-end multitask model that \changed{takes} only the 1D \textsuperscript{1}H and \textsuperscript{13}C NMR spectra and yields a prediction of the molecular structure and substructures (spectrum-to-structure and spectrum-to-substructure, respectively). We achieve a structure accuracy of \numericalresult{60.4\%} within the first 15 predictions for molecules covering the full range of organic elements with up to 40 heavy atoms - a chemical space that contains in excess of \numericalresult{$\sim10^{35}$} molecules. Our framework is thus capable of overcoming the combinatorial scaling of chemical space to make a full prediction of the molecular structure and to rapidly constrain the chemical search space in a wholly unsupervised manner without any context beyond routinely collected 1D NMR spectra. Furthermore, we show that our framework is extensible to experimentally collected NMR spectra, achieving a structure accuracy of \numericalresult{21.5\%} when fine-tuned on only 50 experimental spectra while retaining its accuracy on simulated spectra. Our approach provides a powerful tool for accelerating chemical discovery across the molecular sciences, extending the boundaries of the insights that can be extracted from 1D NMR spectroscopy.

\section{Results and discussion}\label{sec:results_discussion}

\begin{figure}[!ht]
    \centering
    \includegraphics[width=\textwidth]{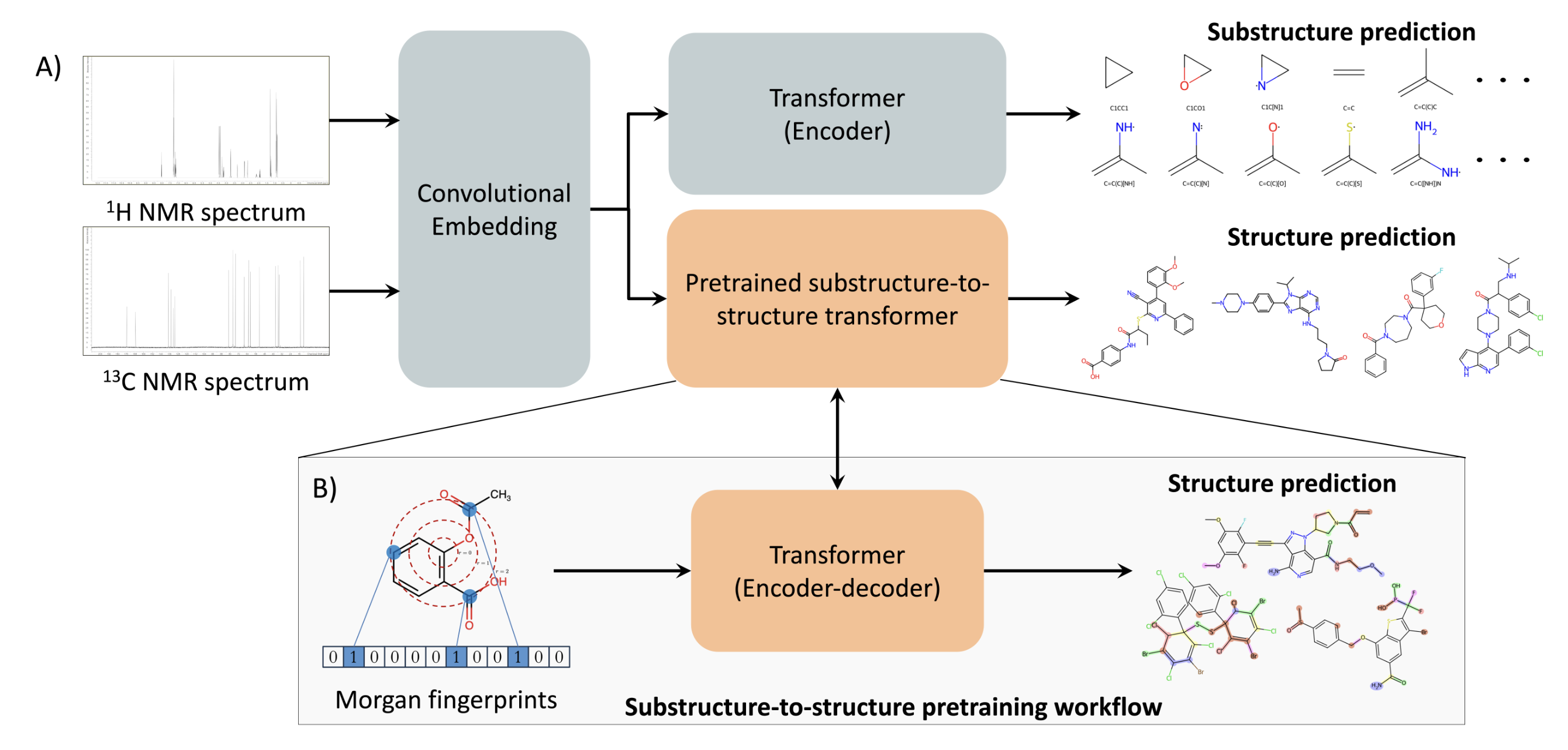}
    \caption{An overview of our structure elucidation framework consisting of (A): the multitask spectrum-to-structure/spectrum-to-substructure model that generates both structure and substructure predictions and (B): the substructure-to-structure pretraining approach that reconstructs SMILES strings from Morgan fingerprints. Weights from a transformer pretrained on the substructure-to-structure task are used to initialize the multitask model, as indicated by the arrow connecting the two and the different coloration of the encoder-decoder component of the multitask framework. Specific details regarding the transformer model architecture and multitask model architecture can be found in SI Section~S1.}
    \label{fig:overview}
\end{figure}

Our overall structure elucidation framework, which takes in 1D (\textsuperscript{1}H and/or \textsuperscript{13}C) NMR spectra and produces predictions for the structure and substructures of a compound, is shown in Figure~\ref{fig:overview}(A). The bottom row (B) outlines our pretraining procedure, where the Morgan fingerprints of molecules are translated into their corresponding SMILES strings, i.e. the substructure-to-structure task. The top row (A) depicts the multitask model which learns to predict the molecular structure and substructures from the \textsuperscript{1}H and/or \textsuperscript{13}C NMR spectra, i.e. the spectrum-to-structure and spectrum-to-substructure tasks, respectively. For the spectrum-to-structure and spectrum-to-substructure tasks our framework uses an end-to-end approach (without any intermediate processing steps) and solely uses the \textsuperscript{1}H NMR and \textsuperscript{13}C NMR spectra without the molecular formula, fragments, or other context. \changed{To initialize the multitask model, we transferred the weights from the pretrained substructure-to-structure model only into the structure elucidation branch of our multitask framework, which is shown in Figure~\ref{fig:overview}(A) by the double-headed arrow. Initializing the substructure elucidation branch from the pretrained model did not lead to improvements in either the substructure prediction performance or structure elucidation performance (see SI Section~S1)}.

The \textsuperscript{1}H NMR spectrum, the Fourier-transformed free induction decay as typically obtained from the spectrometer, is first encoded through a convolutional neural network and \newchanges{this} latent representation is combined \changed{with an embedded representation of the \textsuperscript{13}C NMR spectrum chemical shifts (SI Section~S1)}. This combined representation is then passed to a \newchanges{transformer} encoder-only substructure network, which generates the substructure prediction, and a \newchanges{transformer} encoder-decoder network, which generates the structure prediction (Figure~\ref{fig:overview}(A)). \newchanges{For the substructure elucidation branch in our multitask network, we use a four-layer transformer encoder network that generates a hidden representation from the combined spectrum representation through the repeated application of multithead self-attention and feed-forward layers before passing this hidden representation to a sequence pooling layer\cite{hassani_escaping_2022} to generate the final substructure probabilities. For the structure prediction branch, we use an eight-layer encoder-decoder transformer network that first maps the combined spectrum representation to a hidden memory state and then uses this memory in the decoder to generate a final hidden representation. A linear projection applied to this hidden state then generates a probability distribution over our vocabulary of tokens, which we sample one at a time to generate the SMILES string prediction. We choose this multitask architecture because, unlike the SMILES strings, which are generated one token at a time (i.e., autoregressively), the substructure probabilities from the spectrum are predicted all at once, meaning an encoder network using multithead self-attention is sufficient for capturing the long-range correlations between spectral features that inform this output. For a more detailed description of the different components of our network, see SI Section~S1.}

We first demonstrate the effectiveness of the pretraining procedure on the substructure-to-structure task before showcasing the overall performance of our multitask framework on the spectrum-to-structure and spectrum-to-substructure tasks. \changed{These two tasks produce two complementary outputs that a user might want to learn about their analyte from its NMR spectra: the structure of the molecule, which contains the formula and connectivity information, and the substructures of the molecule, which can provide useful insights in cases where the full structure prediction does not succeed.} \newchanges{Additionally, combining both tasks within one model leads to higher computational efficiency given that we only need to optimize one model as opposed to two separate models.} 

\subsection{Reconstructing molecules from fragments using Morgan fingerprints and transformers}\label{sec:mfp_transformers}
Manual structure elucidation from 1D NMR spectra often proceeds by first deducing the presence of molecular substructures (fragments) based on empirical observations and established chemical shifts and then assembling the identified fragments into the overall molecule. Taking inspiration from this approach, we pretrain our neural network on the substructure-to-structure task, conditioning it via molecular representation learning to understand the semantics and syntactic validity of SMILES string representations of molecules. 

Morgan fingerprints\cite{rogers_extended-connectivity_2010} provide an appealing route for constructing representations of molecular fragments in an element- and size-agnostic way. This enables the rapid construction of compact representations for molecules with significantly greater element and molecule size coverage, allowing for straightforward expansion beyond the sets of substructures covering C, N, O, and H we employed in our previous work\cite{huang_framework_2021,hu_accurate_2024}. Morgan fingerprints represent a molecule as a binary vector. These are constructed by first extracting circular environments of a specified radius around an atom, hashing that extracted environment, and then taking the modulus of the hashed value against the number of bits in the fingerprint to obtain the extracted environment's position within the binary vector. Computing Morgan fingerprints for molecules using RDKit\cite{noauthor_rdkit_nodate} is highly efficient, enabling the rapid assembly of large datasets.

For the substructure-to-structure task (Figure~\ref{fig:overview}(B)), we train a transformer model\cite{vaswani_attention_2017} to translate between the non-zero bits of a molecule's Morgan fingerprint and its SMILES string representation. The only input to the model is a binary vector representing the Morgan fingerprint of the molecule and the output is the SMILES string. For this task, we use Morgan fingerprints with a radius of 2 and a vector size of 8192 bits, which we found provided a good compromise between high reconstruction accuracy and the compactness of the representation (SI Section~S2.1). To train the model, we used all the molecules available (as of February 2025) in PubChem\cite{kim_pubchem_2025} that consisted of up to 40 heavy atoms composed of only the following elements: C, N, O, H, B, P, S, Si, F, Br, Cl, and I. \newchanges{After removing the stereochemical information from the SMILES strings, this yielded a set of 88M unique SMILES strings (SI Section~S2.2)}.

\begin{figure}
    \centering
    \includegraphics[width=\textwidth]{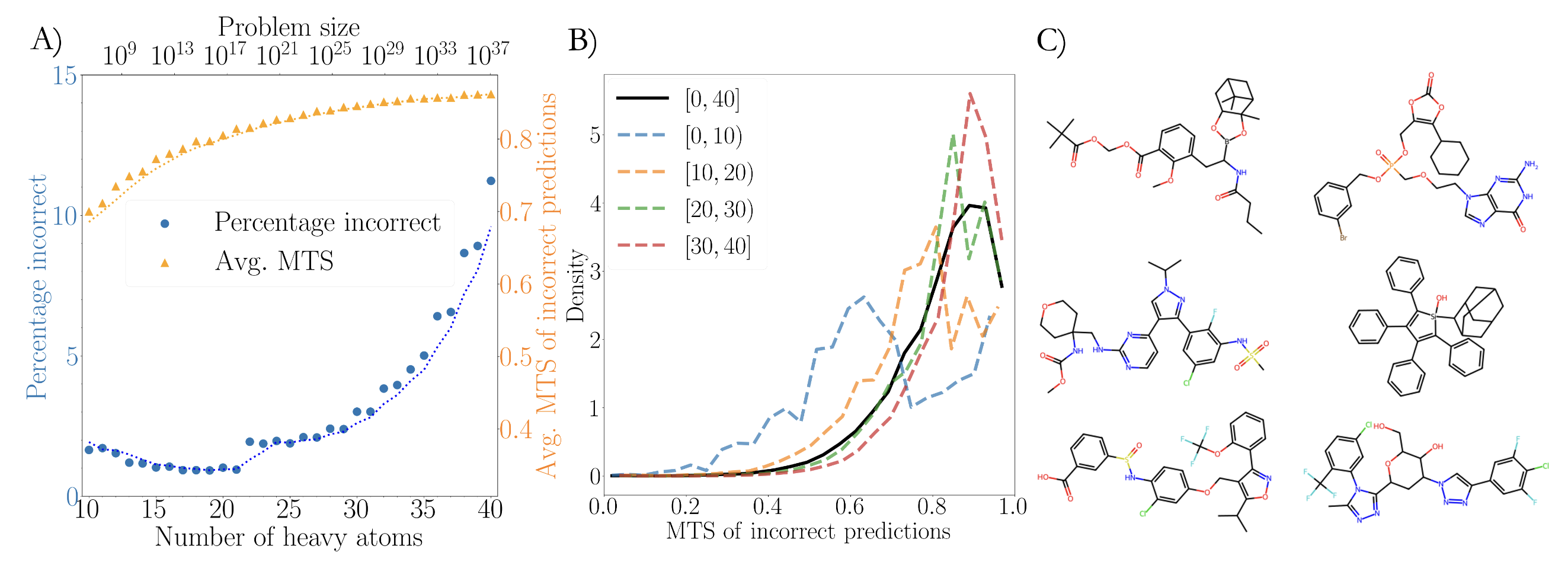}
    \caption{Results for the substructure-to-structure task \changed{using the Top-15 test set structure accuracy}. (A) \changed{Percentage} of incorrect predictions and the average maximum Tanimoto similarity (MTS) of incorrect predictions to the target molecule as a function of the number of heavy atoms. The dashed lines are the moving averages for each quantity across sizes. (B) The distribution of the MTS to the target molecule with decompositions across different ranges of numbers of heavy atoms. (C) Examples of molecules that were predicted correctly by the substructure-to-structure transformer, all with 40 heavy atoms and varied elemental compositions.}
    \label{fig:substruct_trans_results}
\end{figure}

In order to test the performance of our substructure-to-structure model, we generate a set of 15 predictions for each target using the top-$k$ random sampling\cite{fan_hierarchical_2018} procedure during inference time with $k=5$ (SI Section~S2.3). We consider a molecule correctly predicted if it appears within the set of 15 predictions after canonicalization of the model's predictions. We obtain a \changed{Top-15 test set accuracy} of \numericalresult{97.8\%}, \newchanges{across all molecule sizes up to and including 40 heavy atoms,} indicating highly accurate reconstruction of the molecule from its Morgan fingerprint. There is notably only a small decrease in \changed{the Top-15} accuracy of the model as the molecule size increases, with \numericalresult{88.8\%} \changed{Top-15 accuracy} for molecules of 40 heavy atoms (Figure~\ref{fig:substruct_trans_results}(A), blue dots), despite the fact that from 10 to 40 heavy atoms, the estimated chemical space grows by over 30 orders of magnitude (SI Figure~S4). The decrease in accuracy for large compounds likely arises in part from dataset imbalance: PubChem has fewer large molecules ($>$22 heavy atoms) because drug-like compounds are often designed to comply with Lipinski's rule of five\cite{lipinski_lead-_2004}, a set of constraints that encourage oral bioavailability.
Ultimately, only \numericalresult{0.91\%} of the data comes from molecules of 40 heavy atoms in contrast to \numericalresult{6.88\%} from 20 heavy atoms, so the fact that the model's accuracy decays only mildly for systems of this size suggests that the model has learned to \changed{generalize over this complex chemical space}. This generalizability can be attributed to the representational power of Morgan fingerprints: using a radius of 2 and a fingerprint size of 8192, we are able to differentiate all but \numericalresult{138} of the $\sim1.5$M molecules in the test set, indicating that these fingerprints are highly efficient at discriminating between molecules by providing nearly unique sequence representations for each molecule. Figure~\ref{fig:substruct_trans_results}(C) shows examples of molecules that our model correctly predicts in the substructure-to-structure task. These molecules each contain 40 heavy atoms, and their reconstruction requires assembling between \numericalresult{30} and \numericalresult{80} different substructures encoded within their fingerprints, underscoring the complexity of this task. The transformer thus proves to be an effective architecture for learning to invert these unique fingerprints back into the SMILES representation.

It is instructive to also analyze the failures of our substructure-to-structure model. To assess this, for molecules where the exact target was not produced by our substructure-to-structure model we rank the predicted molecules by their Tanimoto similarity\cite{rogers_computer_1960} to the correct target molecule. The maximum Tanimoto similarity (MTS) prediction is then used to assess how close the model got to the correct answer when it does not predict the exact target. The MTS averaged over the molecules of a given size is shown as a function of the molecule size as the orange triangles in Figure~\ref{fig:substruct_trans_results}(A), and the distribution of MTS values observed for different ranges of molecule sizes is shown in Figure~\ref{fig:substruct_trans_results}(B). The average MTS increases with the number of heavy atoms (Figure~\ref{fig:substruct_trans_results}(A), orange triangles). Hence, although the model has a lower accuracy for larger molecules (Figure~\ref{fig:substruct_trans_results}(A), blue dots), the larger MTS indicates that it still recovers a substantial portion of a target molecule's structure from its fingerprint. Average MTS errors in smaller molecules are naturally higher since even a one atom error will likely constitute a larger portion of the molecule's scaffold. Consistent with the increased average MTS for larger molecules, Figure~\ref{fig:substruct_trans_results}(B) shows a rightward shift of the full distribution of MTS values for different ranges of molecule sizes with molecules with less than 10 heavy atoms giving only \numericalresult{79.3\%} predictions above a MTS of 0.5 while for 30 - 40 heavy atoms this increases to \numericalresult{99.2\%} (SI Table~S6). Across all molecule sizes, the model's incorrect predictions have an average MTS of \numericalresult{0.82}, with \numericalresult{97.8\%} of the incorrect predictions exceeding a threshold similarity of 0.5. It should be noted that for Tanimoto similarities $>$0.8, the differences between incorrect predictions and targets are subtle, sometimes differing by as little as a single bond or atom. Some examples of incorrectly predicted molecules and their requisite target molecules for different values of Tanimoto similarity are shown in SI Figure~S5. The model is thus highly accurate in recovering the correct molecular structure across sizes and fails gracefully, often suggesting incorrect predictions that are within the neighborhood of the target molecule.

\subsection{Molecule structure and substructure prediction from spectra using a multitask framework}\label{sec:multitask_perf}
Having established the effectiveness of the substructure-to-structure transformer model, we now integrate this pretrained network into the overall multitask framework for performing the spectrum-to-structure and spectrum-to-substructure tasks. For the spectrum-to-structure and spectrum-to-substructure tasks, our multitask framework \changed{takes} \textsuperscript{1}H and/or \textsuperscript{13}C NMR spectra and produces predictions of both the molecular structure and substructures, as shown in Figure~\ref{fig:overview}(A). The multitask framework consists of an embedding network that creates a joint representation of the \textsuperscript{1}H and/or \textsuperscript{13}C NMR spectra that then feeds into two separate models: a transformer encoder-only substructure elucidation model that predicts the molecule's substructure profile, and an encoder-decoder transformer model that predicts SMILES strings. We find that initializing the structure prediction portion of the multitask framework using weights from the pretrained substructure-to-structure transformer model yields better accuracies in the spectrum-to-structure task compared to training from a random initialization, with an increase in \changed{the Top-15 structure} accuracy of \changed{22 percentage points} (see below). This advantage arises from a combination of the increased data volume available for the substructure-to-structure task (88M vs 2M) compared to the spectrum-to-structure/spectrum-to-substructure tasks and the fact that the molecular representation learned by the transformer during the substructure-to-structure task is highly transferable within the new framework. 

To train our multitask framework, we first selected a subset of 5M molecules from the 88M molecules we used for training the substructure-to-structure transformer, taking care to ensure that there was no data leakage between the training, validation, and test sets used, i.e. preventing molecules that appear in the test set from being part of the training set. Rather than randomly sampling this set of 5M, we use an algorithm based on Morgan fingerprints that biases the selection towards diverse entries for the dataset (SI Section~S3.1, Algorithm~S1). Once we obtained the set of 5M SMILES strings, we then further filtered this set down to 2M molecules using recently introduced algorithms for rapid estimations of molecular similarities and clustering\cite{lopez-perez_isim_2024,perez_bitbirch_2025} (SI Section~S3.1, Algorithm~S2). With this final set of 2M molecules, we forward-simulated the \textsuperscript{1}H NMR spectra and \textsuperscript{13}C NMR \changed{chemical shifts} using the v2024.2 batch NMR predictors from ACD/Labs. We partitioned the spectra using a random split of 80\% for training and 10\% each for validation and testing, again ensuring no leakage of molecules between the sets. 

For the spectrum-to-substructure task we must devise a set of substructures (molecular fragments) that the model can predict for. To accomplish this, one cannot simply predict Morgan fingerprints because while Morgan fingerprints do encode substructure environments, the bits themselves can only be mapped back into substructures if the molecule is known, since the hashing algorithm used to generate Morgan fingerprints is both dependent on the fingerprint size and can be susceptible to hash collisions. Hence, we use the Morgan algorithm to generate a set of consistent substructures, i.e., ones that can be interpreted as individual fragments without knowledge of the corresponding molecule. To do this, we use the Morgan algorithm to enumerate and collate all unique substructures across the entire set of 2M molecules, considering a maximum radius of 1 for the atomic neighborhoods. One should note that a neighborhood can be centered on any atom, so that, for example, chlorine containing substructures include multiple ones originating on carbon such as \texttt{[C]C([S])Cl} where a central carbon is bonded to a carbon, sulfur, and chlorine; \texttt{[CH]C(Cl)(Cl)Cl} where a central carbon is bonded to three chlorines and one carbon; and \texttt{[CH]C(=O)Cl} where a central carbon is double bonded to an oxygen, a chlorine, and another carbon. After filtering out exceedingly rare substructures that have a \changed{maximum fraction of occurrence across the set of 2M molecules lower than \numericalresult{$2.5\times 10^{-5}$}}, this results in the set of $\sim$3K substructures that our model predicts (SI Section~S3.2).

\begin{figure}
    \centering
    \includegraphics[width=\textwidth, height=0.33\textheight, keepaspectratio]{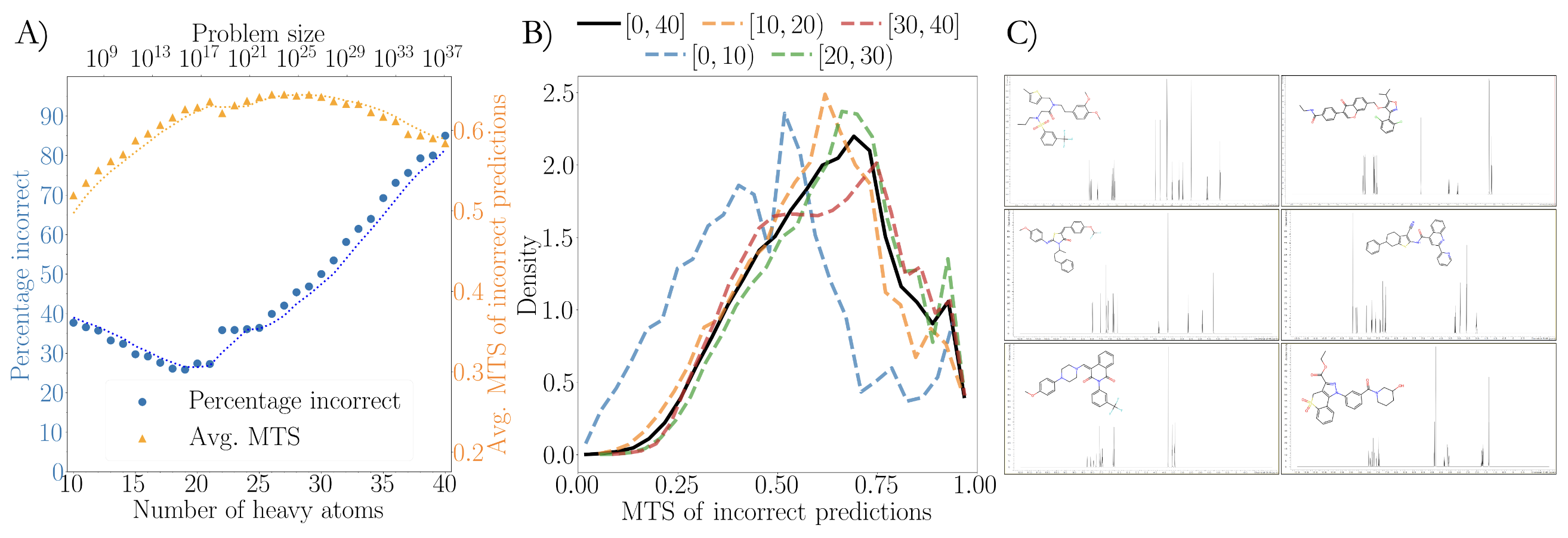}
    \caption{Results for the spectrum-to-structure task \changed{using the Top-15 test set structure accuracy}. (A) \changed{Percentage} of incorrect predictions and the average maximum Tanimoto similarity (MTS) of incorrect predictions to the target molecule as a function of the number of heavy atoms. The dashed lines are the moving averages for each quantity across sizes. (B) The distribution of the MTS to the target molecule with decompositions across different ranges of numbers of heavy atoms. (C) Examples of molecules that were predicted correctly by the multitask model alongside their \textsuperscript{1}H NMR spectra. All systems shown range from 35 - 40 heavy atoms.}
    \label{fig:spec_to_struct_results}
\end{figure}

To assess our multitask framework, we first consider its performance on the spectrum-to-structure task. Similar to our metrics for evaluating the substructure-to-structure transformer, we sample a set of 15 structures from the multitask model using top-$k$ random sampling with $k=5$ and consider a target string correctly predicted if it appears within the set of predictions after canonicalization. Across the entire test set, we obtain \changed{a Top-15 structure accuracy} of \numericalresult{60.4\%}, with the model retaining predictive power across molecules ranging in size from 10 to 40 heavy atoms. This is remarkable as the model achieves this accuracy using only the \textsuperscript{1}H and \textsuperscript{13}C NMR spectra despite the chemical space growing by over 30 orders of magnitude over this range of molecule sizes. Figure~\ref{fig:spec_to_struct_results}(A) shows that accuracy of the model declines from around \numericalresult{73\%} for systems of 20 heavy atoms to around \numericalresult{15\%} for systems of 40 heavy atoms, a decrease in accuracy that is small when compared to the growth in the size of chemical space from an estimated \changed{$10^{15}$ molecules at 20 heavy atoms to $10^{37}$ at 40 heavy atoms, an increase of 22 orders of magnitude.} Similar to our observation in the substructure-to-structure task, a more rapid decline in accuracy is seen for system sizes greater than 22 heavy atoms due to the increased rarity of these molecules in PubChem, though its effect is more pronounced here. This suggests that the model's accuracy for larger systems could be improved by reducing the dataset imbalance, in which larger systems are proportionally underrepresented. \newchanges{We note that, unlike in the substructure-to-structure case, the accuracy of the multitask model noticeably declines for systems of less than 20 heavy atoms. This non-monotonic behavior likely arises from the non-uniform distribution of molecular sizes in the training dataset (SI Figure~S6) combined with the increased difficulty of the spectrum-to-structure task and the smaller overall volume of data, resulting in a more pronounced decrease in performance. This suggests that accuracy on smaller systems could potentially be improved by including more small molecules in the training set.}

To underscore the complexity of the molecules being predicted using solely their \textsuperscript{1}H and \textsuperscript{13}C NMR spectra, Figure~\ref{fig:spec_to_struct_results}(C) shows molecules from the test set that the model correctly predicted alongside their \textsuperscript{1}H NMR spectra. Our new model is also a marked improvement on our previous generation\cite{hu_accurate_2024}, which only predicted for systems containing C, N, O, and H. Using the test set from this previous work, consisting of systems with only C, N, O, and H with up to 19 heavy atoms, our new model achieves \changed{a Top-15 accuracy} of \numericalresult{77.5\%} compared to a previously reported 69.6\%. This shows that despite increasing the scope of the chemistry that the model is exposed to during training---increasing the elemental \changed{coverage} to also include B, P, S, Si, F, Br, Cl, and I and expanding the maximum molecule sizes it is trained on from 19 to 40---it has the capacity to learn correlations between NMR signals and structures for significantly more complicated systems without sacrificing performance on smaller systems.

\begin{figure}
    \centering
    \includegraphics[width=\textwidth, height=0.33\textheight, keepaspectratio]{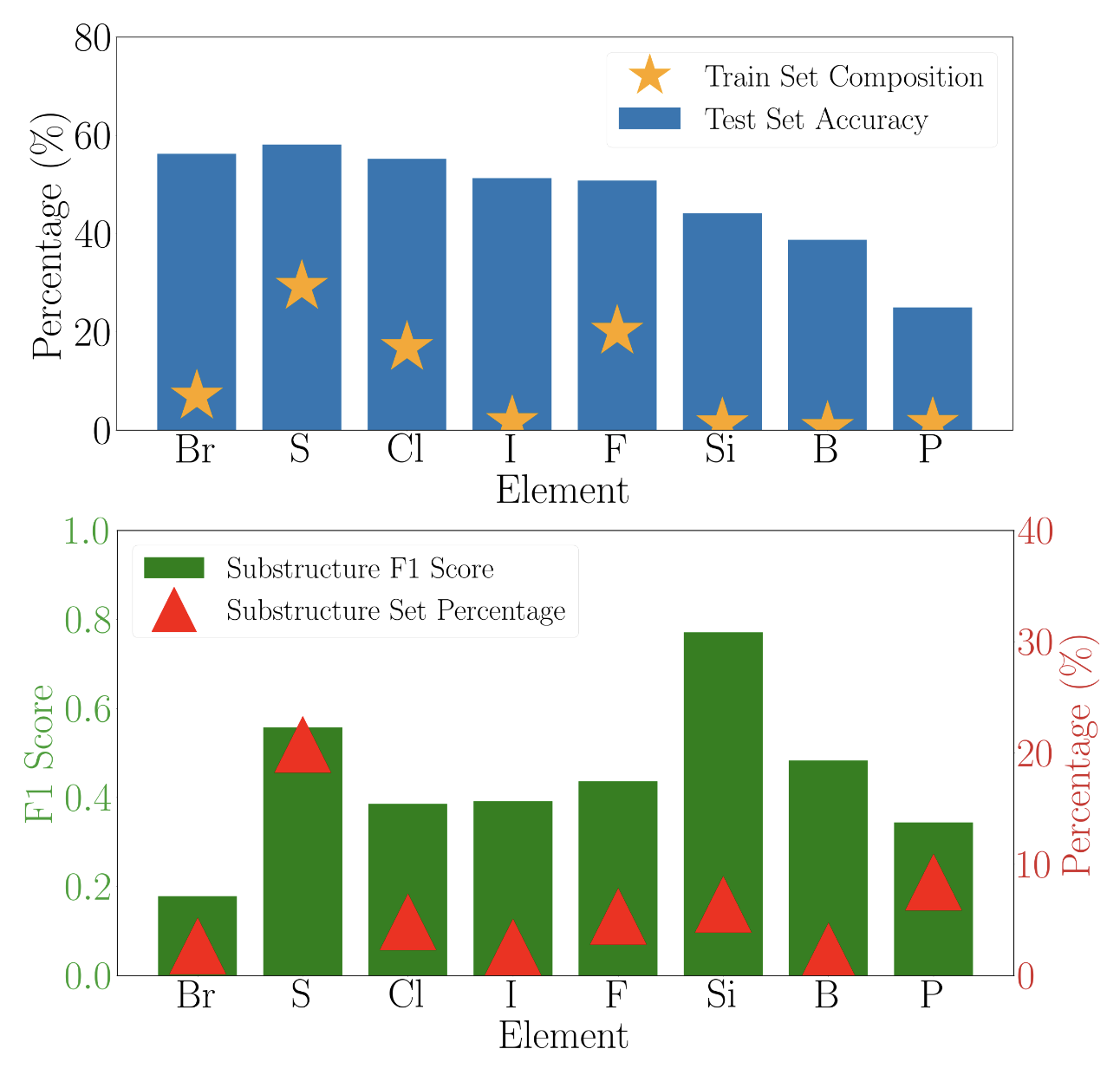}
    \caption{(Top) Multitask model \changed{Top-15} test set structure prediction accuracy resolved by element, where accuracy is the percentage of systems containing that element that the model predicted correctly. (Bottom) Test set substructure prediction accuracy resolved by element, where the accuracy is expressed as the F1 score and the substructure set \changed{percentage} is the \changed{percentage} of substructures in the entire substructure set that contain the specified element.}
    \label{fig:multitask_results_across_elems}
\end{figure}

We can also quantify how our multitask framework's proficiency on the spectrum-to-structure task depends on the elements in the test molecule. The top panel of Figure~\ref{fig:multitask_results_across_elems} shows the \changed{Top-15} test set accuracy of our model's predictions resolved by whether a molecule contains a particular element. \changed{From this data, one sees that the model has learned to generalize and correctly predict the presence of atoms beyond C, N, and O, even in cases where systems containing those elements are very rare in the training data. For example, I, Si, B, and P each occur in less than 2\% of the training set molecules, and B in less than 1\%, but all have accuracies of $>20\%$. In general, predictions for polyvalent elements are more challenging than the monovalent halides. Sulfur is an exception with the highest prediction accuracy \numericalresult{(58.1\%)}, which is consistent with its higher training set composition, shown as stars in Figure~\ref{fig:multitask_results_across_elems}, with S occurring in \numericalresult{29.3\%} of molecules. Interestingly, the prediction accuracy is lower for P than Si or B despite its higher occurrence in the training set than these atoms. This result may arise from the greater variety of ways in which P is incorporated into molecular scaffolds, split between instances in which P has a valency of 5 (\numericalresult{63\%}) or 3 (\numericalresult{37\%}), while Si and B have valencies of 4 and 3, respectively. (Boron can have a valency of 4 in, for example, anionic borate compounds, but the B-containing compounds in our dataset are all valency 3).}   This greater flexibility when bonding to P is corroborated by the \numericalresult{237} substructures containing P as opposed to \numericalresult{181} for Si and \numericalresult{62} for B. The challenges in the model's generalization to P are further reaffirmed by resolving the performance of the substructure-to-structure transformer based on the elemental composition (SI Figure~S7), which also shows the model's worse performance on P-containing systems relative to Si- and B-containing systems, despite the higher frequency of occurrence for P-containing systems. Given that the substructure-to-structure transformer is tasked with reconstructing molecules from representations of their substructure environments, this reinforces the idea that the model is learning to generalize between elements of similar valencies. For Si and B with valencies of 4 and 3, respectively, the model learns to predict the presence of these atoms by training on their analogues, C and N, which are abundant in the dataset. However, due to P's mixed valency and the resulting increased diversity of environments, the model generalizes less effectively, leading to lower accuracy in systems containing P relative to Si and B.

To analyze the substructure performance obtained from the spectrum-to-substructure task of our multitask framework, we must account for the fact that each individual molecule contains only a small portion of the total set of substructures (on average \numericalresult{2.5\%}), leading to a naturally imbalanced classification problem with far more true negatives than true positives. Defining accuracy using only true positives and true negatives is thus a poor metric in this scenario, because even a naive predictor that always predicts that no substructures are present in a molecule will spuriously achieve high accuracy. To counteract this, we use the F1 score, a metric that balances the precision and recall of the model. As shown in the first row of Table~\ref{table:spectrum_ablations}, the F1 score of our model when using both the \textsuperscript{1}H and \textsuperscript{13}C NMR spectra is \numericalresult{0.84}. On the narrower (C, N, O, H only and $<$20 heavy atoms) test set from our previous generation model\cite{hu_accurate_2024} the new model achieves the same F1 score, 0.86, as the previous generation model, emphasizing that expanding the model to larger molecules and a wider range of elements does not compromise its performance on simpler systems. The output of the spectrum-to-substructure task is a vector of probabilities $\mathbf{p}$ where each $p_{i}\in \mathbf{p}$ represents the probability that the model predicts for the substructure being present in the system from which the spectrum was obtained. On the test set, only \numericalresult{1.9\%} of the substructure predictions from our model are within the range of $0.1 < p_i < 0.9$, i.e., \numericalresult{98.1\%} are either $> 0.9$ or $< 0.1$, indicating that the model typically has high confidence in its predictions. When the model predicts a substructure being present, using a decision boundary of 0.5 between present and absent, with a probability $p_i < 0.1$, it is \numericalresult{99.8\%} accurate in identifying a true negative, and when the model predicts a substructure with a probability $p_i > 0.9$, it is \numericalresult{96.2\%} accurate in identifying a true positive (SI Figure~S13). Resolving the substructure F1 score by element (Figure~\ref{fig:multitask_results_across_elems} bottom) one observes that for all substructures containing a non C, N, O atom the F1 score is lower than the average of \numericalresult{0.84} with Si giving the highest value (\numericalresult{0.77}) and Br the lowest (\numericalresult{0.18}). \newchanges{A possible explanation for why Br containing substructures are a particular challenge (in addition to the fact that they occur as only \numericalresult{2.7\%} of all substructures) is that the range of chemical shifts observed for protons adjacent to Br and Cl overlap. These similar chemical shifts likely lead to a greater likelihood of the model confusing Br with more prevalent monovalent halogens in the training set such as Cl and F, resulting in higher model uncertainty when assigning Br containing substructures. Indeed, in} systems that only contain Br, we find Cl and F containing substructures are predicted \numericalresult{24.4\%} of the time. By contrast, Si substructures have a high F1 score (\numericalresult{0.77}) despite occurring in only \numericalresult{6.4\%} of all substructures, which is likely due to the distinctive chemical shifts of protons in moieties containing Si. 

In some cases, performing \textsuperscript{13}C NMR in addition to \textsuperscript{1}H NMR can be prohibitive. Hence, it is informative to examine how the model's performance changes when only \textsuperscript{1}H NMR is used. Training a spectrum-to-structure model that only takes the \textsuperscript{1}H spectra, the model's \changed{Top-15} structure prediction accuracy drops from \numericalresult{60.4\%} to \numericalresult{46.6\%}, while its substructure F1 score drops from $0.84$ to $0.81$ (Table~\ref{table:spectrum_ablations}). This relatively mild performance degradation is consistent across molecule size (SI Figure~S14) and indicates that the \textsuperscript{1}H NMR spectrum contains the majority of the connectivity information required for elucidating the molecular structure, with the \textsuperscript{13}C NMR providing a small amount of clarifying information in some cases. Indeed, when the model is trained on only the \textsuperscript{13}C NMR, the \changed{Top-15} accuracy drops dramatically to only \numericalresult{19.4\%}, with \changed{low} predictive power \numericalresult{($<$ 10\%)} retained for system sizes above 33 heavy atoms (SI Figure~S14). 

\begin{table}[!ht]
    \centering
    \caption{\changed{Performance of the multitask framework for the spectrum-to-structure and spectrum-to-substructure tasks with different input spectra. Top-N structure accuracies for Top-1 through Top-30 are shown}}
    \label{table:spectrum_ablations}
    \sisetup{round-mode=places}
    \large
    \resizebox{\columnwidth}{!}{\begin{tabular}{ccccccS[round-precision=1]c}
        \toprule
        \textbf{Spectra used} & \textbf{Pretrained} & \textbf{Top-1} & \textbf{Top-5} & \textbf{Top-10} & 
        \textbf{Top-15} & \textbf{Top-30} &
        \textbf{Substructure} \\ 
        \textbf{} & 
        \textbf{transformer} & 
        \textbf{accuracy (\%)} &
        \textbf{accuracy (\%)} &
        \textbf{accuracy (\%)} & 
        \textbf{accuracy (\%)} & 
        \textbf{accuracy (\%)} & 
        \textbf{F1 score} \\
        \midrule
        \textsuperscript{1}H and \textsuperscript{13}C & Yes & \numericalresult{38.9} & \numericalresult{53.1} & \numericalresult{57.9} & \numericalresult{60.4} & \numericalresult{64.3} & \numericalresult{0.84}\\ 
        \textsuperscript{1}H only & Yes & \numericalresult{25.4} & \numericalresult{38.9} & \numericalresult{43.9} & \numericalresult{46.6} & \numericalresult{51.0} & \numericalresult{0.81}\\ 
        \textsuperscript{13}C only & Yes & \numericalresult{4.5} & \numericalresult{12.2} & \numericalresult{16.6} & \numericalresult{19.4} & \numericalresult{24.1} & \numericalresult{0.70}\\ 
        \textsuperscript{1}H and \textsuperscript{13}C & No & \numericalresult{18.9} & \numericalresult{31.1} & \numericalresult{35.8} & \numericalresult{38.4} & \numericalresult{42.6} & \numericalresult{0.84}\\  \midrule
    \end{tabular}}
\end{table}

Table~\ref{table:spectrum_ablations} shows that pretraining has a significant positive influence on the structure accuracy of the model, with an increase of \changed{22 percentage points} in the Top-15 accuracy when using the pretrained transformer as opposed to no pretraining. This suggests that the molecular representation learned by the transformer during the substructure-to-structure task is an effective way of conditioning the model with an understanding of molecular structure that transfers well to the spectrum-to-structure task. \changed{In contrast, the pre-training procedure does not affect substructure prediction performance with the model obtaining comparable F1 scores with (\numericalresult{0.84}) and without (\numericalresult{0.84}) the pre-trained transformer.} This is due to the separation within the architecture: the encoder-only model for predicting the substructure profile does not share any weights with the encoder-decoder transformer used for predicting the structure, with the two models operating largely independently while sharing a joint representation of the spectra as input. 

\changed{Comparing the changes upon moving from Top-1 to Top-30 sampling in Table~\ref{table:spectrum_ablations}, one sees that the structure accuracy of the model increases as more structures are sampled. However, this increase in accuracy quickly diminishes, with improvements beyond Top-15 accuracy becoming increasingly marginal: doubling the sampling from Top-15 to Top-30 increases the accuracy by less than \numericalresult{5\%} in all cases.} \newchanges{Further analysis of the Top-N accuracies, along with descriptive statistics on the position of the correct molecules, can be found in SI Section~S3.5.}

\changed{In addition to assessing whether the correct prediction is among any of the model's first Top-N predictions, one can also assess the model's Rank-1-of-N performance, i.e., its ability to identify the correct candidate from its generated candidate pool. Rank-1-of-N performance is potentially useful in cases where the user has no other knowledge of which molecule is correct and thus wants the model itself to provide a decision on which of the N molecules is most likely to be the correct structure. There are two metrics we consider to address this: (i.) using the substructure predictions to rank the N candidates and (ii.) using the structure prediction transformer's own probability distribution (SI Section~S3.5).} 

\changed{The former of these gives noticeably worse performance as N increases, with Rank 1-of-5 accuracy of \numericalresult{42.2\%} and only \numericalresult{37.9\%} for Rank 1-of-30 (SI Table~S8 top), indicating that as more candidates are generated the model is predicting more incorrect structures that contain substructures that are more consistent with its substructure prediction, making it more difficult to rank the correct structure as the most likely to be correct. However, the latter approach, where the ranking of the candidates is performed using the structure predictor's probability distribution metric, shows systematic improvement in the accuracy as N increases with Rank 1-of-5 accuracy of \numericalresult{47.0\%} and Rank 1-of-30 of \numericalresult{48.6\%} (SI Table~S8 bottom).}

\changed{The ratio of the Rank-1-of-N accuracy (SI Table~S8) to the Top-N accuracy (Table~\ref{table:spectrum_ablations}) for a given N gives the ranking accuracy, since if the ranking metric was exact then it would give a ratio of 1 since it would always select the correct molecule from the N predictions. For example, the Rank-1-of-15 accuracy using the structure predictor's probability distribution is \numericalresult{48.3\%} whereas its Top-15 accuracy is \numericalresult{60.4\%} and thus the structure predictor ranking metric is accurate \numericalresult{80.0\%} of the time in predicting which of the 15 predictions is the correct structure. This ranking metric can therefore be used to reliably obtain correct Rank-1 predictions in cases where leveraging additional information, such as forward simulating and matching the spectra or the user's own chemical insights, is not feasible.}

To understand the failure cases of our multitask model on the spectrum-to-structure task we consider the average MTS for systems where using the model did not predict the correct structure in its predictions from their \textsuperscript{1}H NMR and \textsuperscript{13}C NMR spectra. Across all molecule sizes, the MTS is \numericalresult{0.62} for the spectrum-to-structure model compared to \numericalresult{0.82} for the substructure-to-structure model, emphasizing the significantly greater difficulty of this task. The orange points in Figure~\ref{fig:spec_to_struct_results}(A) show the average MTS value as a function of the molecule size where, in contrast to the substructure-to-structure task (Figure~\ref{fig:substruct_trans_results}(A)), it decreases above 30 heavy atoms, indicating that the model's incorrect predictions exhibit larger structural deviations from the target structure. The likely reason for this is that, as the size of the molecules increases, their NMR spectra become more congested with peaks (see, for example, the spectra in Figure~\ref{fig:spec_to_struct_results}(C)), making the spectrum-to-structure task even more challenging than the substructure-to-structure task as the molecules get larger. To test this hypothesis, one can examine how the F1 score of the substructure predictions changes as the number of heavy atoms (SI Figure~S15) increases, which provides an indication of the challenges the model faces in extracting individual molecular fragments from the spectral data. From this, one sees that for molecules larger than 30 heavy atoms the F1 score decreases, reinforcing the hypothesis that the model is struggling to extract and synthesize relevant structural information from the NMR spectrum. Despite this increased difficulty, \numericalresult{72.2\%} of the incorrect predictions remain above a similarity of 0.5. Taking the test set molecules from our previous generation model\cite{hu_accurate_2024}, which contains smaller molecules and only C, N, O and H, and using their ACD/Labs predicted NMR spectra as input, we observe that our new model's performance on the incorrectly predicted molecules is improved, with an average MTS value of \numericalresult{0.65} compared to 0.56. This further reinforces the fact that scaling the model to larger systems has not come at the cost of performance on smaller systems.

By resolving the MTS distributions by the number of heavy atoms (Figure~\ref{fig:spec_to_struct_results}(B)), we see that as one progresses from systems with $0 \leq n < 10$ heavy atoms (blue dashed line) to systems with $20 \leq n < 30$ heavy atoms (green dashed line), the distributions systematically shift rightward towards higher MTS values, reflecting the decreasing influence of any individual error within larger molecular scaffolds similar to what is observed for the substructure-to-structure transformer in Figure~\ref{fig:substruct_trans_results}(B). However, for systems in the range of $30 \leq n < 40$ heavy atoms (red dashed line), although the distribution's maximum is right-shifted relative to those for systems with less than 20 heavy atoms, it has lower density in the higher MTS region than the one for $20 \leq n < 30$, indicating the presence of greater structural deviations from the target that is consistent with the trend seen for the average MTS value in Figure~\ref{fig:spec_to_struct_results}(A). 

The success of our approach in what seems like a formally insurmountable ($\mathcal{O}(10^{60})$) chemical space naturally reflects that chemistry, and perhaps even more so databases such as PubChem, have extensive correlations in the molecules they contain based on synthesizability, reagent availability, and their chemical utility, which the model is learning to exploit. One can therefore consider an alternative way to assess our model as a generator of candidate molecules, even when it fails to make the correct prediction, and ask the question: is the molecule the model generates in those cases from just seeing the 1D NMR spectra a closer match to the correct molecule than anything that can be found in its PubChem training set? This is a particularly harsh test on our model, since it is not shown the target molecule but only its NMR spectra, yet in this test our model is assessed on whether it does better at generating a molecule that is closer than simply systematically searching the PubChem training set with full knowledge of the target molecule. This test naturally accounts for the extensive correlations in chemical space that we noted above and, accordingly, makes it harder for our model. Specifically, if there is a molecule in the training set that is very similar to a test set molecule, then our model might be expected to do relatively well at predicting the correct structure since it has seen something similar. However, since that molecule is in the test set, to ``win" in this test our model has to predict something closer to the correct test set molecule than the molecule it was trained on. \newchanges{This test therefore directly assesses the competency of our model as a molecular generator in cases where the exact structure is not recovered.} 

\newchanges{To perform this evaluation, we collected all molecules with 40 heavy atoms in our test set that the model was unable to correctly predict within its Top-15 predictions. For each failed test set molecule, we computed the Tanimoto similarity between each of the model’s incorrect predictions and the target, retaining the maximum similarity obtained. We then compute the Tanimoto similarities of every 40 heavy atom molecule in the PubChem training set against the target, where these similarities are sorted in descending order. We then find for each target the first position where the model’s maximum Tanimoto similarity exceeds the sorted PubChem similarities. In this context, a Top-1 position for the model’s prediction means that its best incorrect prediction is closer to the target than any PubChem training molecule considered; a Top-2 position means the best incorrect prediction is the second closest to the target compared to all of the training molecules, and so on. We can then measure the model’s ability to generate close-to-target molecules by the percentage of molecules encapsulated by a given threshold, i.e. the percentage with position less than or equal to the Top-N threshold for a given N, with the objective being to capture the largest percentage possible with the lowest threshold.}  

\newchanges{Despite this tough test, our model shows itself to be a highly capable generator of candidate molecules. For systems of 40 heavy atoms where the model fails to predict the correct molecule, the model’s best incorrect prediction is in the Top-1 position based on Tanimoto similarity to the target \numericalresult{32.2\%} of the time when compared to the best possible match from all 40 heavy atom systems in the PubChem training dataset of 85M molecules (of which 779776 contain 40 heavy atoms). From here, the percentage of molecules captured by the Top-N threshold rises steadily for N from 1 to 30, with the model’s prediction among the top 30 candidates for all molecules in the PubChem training dataset \numericalresult{88.3\%} of the time (SI Figure~S16)}. Combined with the fact that the model is very fast when generating predictions \changed{(\numericalresult{3.39} seconds for a single prediction using an AMD EPYC 9354 32-Core CPU or \numericalresult{0.64} seconds on a single H100 GPU)}, this indicates that even when the model is not capable of generating the correct prediction, it can be used to rapidly constrain the chemical search space to candidate molecules that are closer to the target system when compared to brute force dataset searches.

\newchanges{To increase the utility of models for structure elucidation from NMR spectroscopy for real-life laboratory applications, it is necessary to train the model with experimental data to better capture the natural variability of real measurements. To achieve this, the current sparsity of publicly available high-quality, well-characterized experimental NMR data produced at the instrument's data resolution must ultimately be addressed.} For example, sorting the Biological Magnetic Resonance Data Bank (BMRB)\cite{hoch_biological_2023} for compounds with \textsuperscript{1}H and \textsuperscript{13}C spectra collected on 400 - 500 MHz spectrometers in CDCl$_3$ yields only 100 unique spectra as of December, 2025. However, having been trained on a large collection of simulated spectra, the models we introduce in this paper have learned a chemically-informed latent space conditioned on NMR and thus could be viewed as foundational models that can be specialized for specific uses. As a demonstration of this, we performed supervised fine-tuning of our model using experimental spectra, with 50 of the 100 BMRB spectra for training, 25 for validation, and 25 for testing. Despite this exceptionally small fine-tuning training set, we achieve an average \changed{Top-15} structure accuracy of \numericalresult{21.5\%} on the experimental test set (see SI Section~S3.6), while retaining essentially all of the model's performance on the simulated set (structure accuracy of \numericalresult{60.2\%} vs \numericalresult{60.4\%} before fine tuning). \newchanges{This is a dramatic improvement over the zero-shot accuracy of the model on the experimental test set, which is \numericalresult{0.0\%} owing to the high degree of specialization of the initial model to the simulated data. This ability to successfully tune our model even in the very small data regime again reinforces their potential as foundational structure elucidation models, so to facilitate their use in this regard, we have released our models and the data used to train them on \changed{Zenodo\cite{hu_datasets_2026}} as well as the code as an open-source repository on \changed{GitHub\cite{hu_nmrelucidator_nodate}}.}

\section{Conclusion}

In summary, we have introduced a multitask framework capable of elucidating the structure and substructures of molecules with up to 40 heavy (non-hydrogen) atoms with an elemental coverage spanning the elements most commonly encountered in organic chemistry: C, N, O, H, P, S, Si, B, F, Cl, Br, and I. We showed that our pretraining procedure allows us to reconstruct molecules in the PubChem dataset from their Morgan fingerprints using a transformer architecture, achieving \numericalresult{97.8\%} accuracy within the first 15 predictions. We then integrated our pretrained transformer model into a multitask framework to predict both molecular structure and substructures from only \textsuperscript{1}H and \textsuperscript{13}C NMR spectra with minimal preprocessing, achieving \numericalresult{60.4\%} structure accuracy on the task of \textit{de novo} structure generation within the first 15 predictions. Remarkably, our model is also capable of achieving much of its predictive accuracy using only the \textsuperscript{1}H NMR spectra, with a \changed{Top-15} structure accuracy of \numericalresult{46.6\%}, indicating its ability to extract the rich information encoded in NMR spectra. Furthermore, we demonstrated that the model's predictions of substructures are highly accurate, with \numericalresult{98.1\%} of substructure predictions falling within the region of high confidence ($p_i > 0.9$ or $p_i < 0.1$) and the model predicting true negatives and true positives with accuracies of \numericalresult{99.8\%} and \numericalresult{96.2\%}, respectively. Thus, our framework not only overcomes the combinatorial scaling of chemical space (estimated to be as large as $10^{60}$ for systems of up to 40 heavy atoms) using nothing else besides the 1D NMR spectra, but also provides additional insight into the local chemical environments that appear for a given spectrum which can be used to further guide and accelerate the structure elucidation process. Furthermore, this technology could be used in a synergistic manner with other structure elucidation frameworks, such as CASE, to create new, highly flexible tools for structure elucidation.

In addition to its utility as a standalone tool, we envision many powerful uses for our framework in the medium term in conjunction with other technologies. For example, our framework could be integrated to seed more exhaustive search-based structure elucidation methods with reasonable candidates, or to serve as a platform for reaction characterization and monitoring by injecting additional information into the model's context. While additional challenges remain to achieve full \textit{de novo} structure elucidation from 1D NMR spectra (e.g. stereochemical determination and bridging the gap between simulated and experimental data), we envision that our multitask framework and approach can serve as a foundation for moving towards fully automated, large-scale structure elucidation.

\begin{suppinfo}
The supporting information contains the architectures and hyperparameters for all models presented in the main text, details on the data curation for the substructure-to-structure, spectrum-to-structure, and spectrum-to-substructure tasks, optimization protocols, additional analyses for all models presented in the main text, and additional examples of molecules with the associated spectra or substructures. \thirdchanges{Additional references related to the machine learning techniques and datasets used are included in the SI\cite{schwaller_found_2018,paszke_pytorch_2019,hinton_improving_2012,ba_layer_2016,fink_virtual_2007,blum_970_2009,ruddigkeit_enumeration_2012,bremser_hose_1978,willcott_mestre_2009,bemis_properties_1996}.}
\end{suppinfo}

\begin{acknowledgement}
F.H. acknowledges support from a Stanford Center for Molecular Analysis and Design (CMAD) fellowship. Some of the computing for this project was performed on the Sherlock cluster. We would like to thank Stanford University and the Stanford Research Computing Center for providing computational resources and support that contributed to these research results. J.M.T acknowledges support from the Stanford Undergraduate Research Fellowship in Chemistry. We would like to thank Hongyi Zhang for helpful discussions on NMR spectroscopy.
\end{acknowledgement}

\section{Data and Software Availability}
All code for training and evaluating the models presented in this work can be found on GitHub at https://github.com/MarklandGroup/NMRElucidator. The data and scripts used to produce the datasets can be found on Zenodo at https://doi.org/10.5281/zenodo.19239796.

\section{\fourthchanges{Author Contributions}}

\fourthchanges{F.H.: Conceptualization, Data Curation, Formal Analysis, Investigation, Methodology, Project Administration, Resources, Software, Supervision, Validation, Visualization, Writing - original draft, Writing - review and editing. J.M.T.: Data Curation, Formal Analysis, Investigation, Methodology, Software, Validation, Visualization. D.A.: Data Curation, Investigation, Resources, Software, Validation, Visualization, Writing - review and editing. S.G.: Data Curation, Investigation, Resources, Software, Validation, Visualization, Writing - review and editing. M.E.: Data Curation, Investigation, Resources, Software, Validation, Visualization, Writing - review and editing. G.M.R.: Conceptualization, Investigation, Methodology, Project Administration, Resources, Software, Supervision, Validation, Visualization, Writing - original draft, Writing - review and editing. M.W.K.: Conceptualization, Investigation, Methodology, Project Administration, Resources, Supervision, Validation, Visualization, Writing - original draft, Writing - review and editing. T.E.M.: Conceptualization, Funding Acquisition, Investigation, Methodology, Project Administration, Resources, Supervision, Validation, Visualization, Writing - original draft, Writing - review and editing. All authors have given approval to the final version of the manuscript.}

\newpage
\bibliography{references}

\clearpage
\noindent For Table of Contents Only:

\includegraphics[width=1.0\columnwidth]{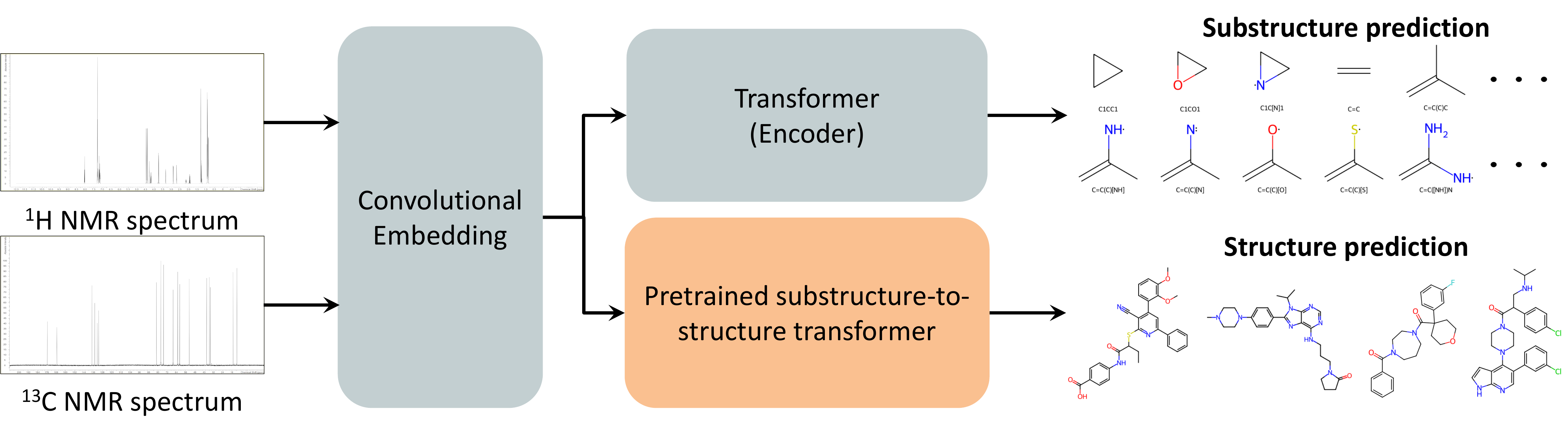}

\end{document}


\maketitle
\tableofcontents
\newpage

\section{Details on model architecture}\label{SI_sec:model_arch}

\begin{figure}[!h]
    \centering
    \includegraphics[width=\textwidth]{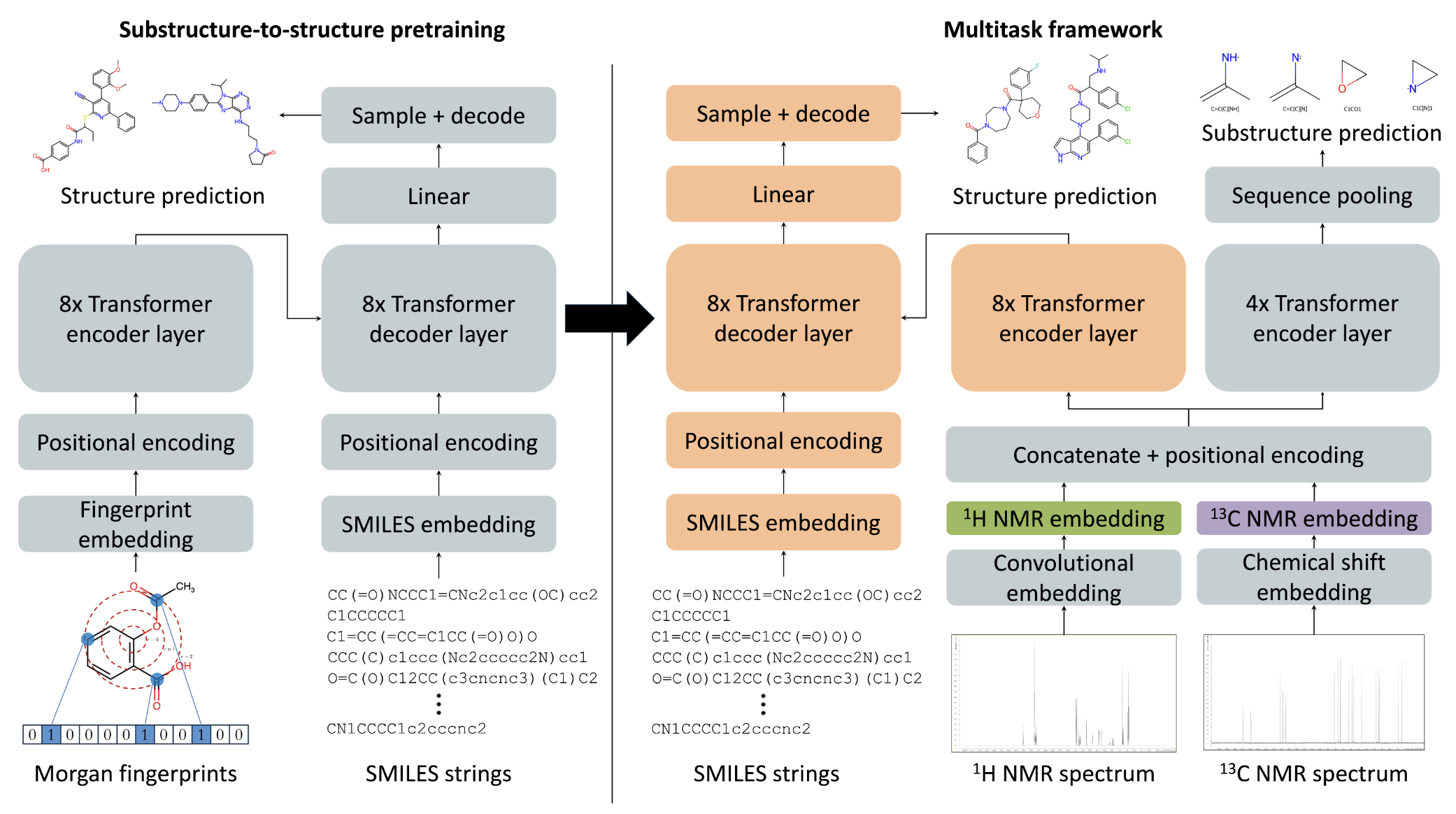}
    \caption{An overview of our structure elucidation framework consisting of (right): the multitask spectrum-to-structure/spectrum-to-substructure model that generates both structure and substructure predictions and (left): the substructure-to-structure pretraining approach that reconstructs SMILES strings from Morgan fingerprints. Weights from a transformer pretrained on the substructure-to-structure task are used to initialize the multitask model, as indicated by the arrow connecting the two and the different coloration of the encoder-decoder component of the multitask framework.}
    \label{SI_fig:overview}
\end{figure}

Our framework for automated structure elucidation from NMR spectra consists of a two-step protocol, as shown in Figure~\ref{SI_fig:overview}. We first train a substructure-to-structure transformer model on the task of reconstructing the molecular structure from the associated Morgan fingerprint (Figure~\ref{SI_fig:overview}(left)) before using those weights to initialize the multitask framework for the spectrum-to-structure and spectrum-to-substructure tasks (Figure~\ref{SI_fig:overview}(right)). The transformers that we use are based on the original encoder-decoder architecture presented in Vaswani et al.\cite{vaswani_attention_2017}, and sinusoidal positional encodings are used throughout.

For the substructure-to-structure transformer model (Figure~\ref{SI_fig:overview}(left)), Morgan fingerprints\cite{rogers_extended-connectivity_2010} are used as the representation of the molecule and are provided to the transformer as a binary vector. This binary vector is then transformed into a right zero-padded sequence of one-indexed integers corresponding to the positions of the on bits. This enables direct embedding of the bit positions and also leads to a more compact representation since the number of bits typically populated for any given molecule is significantly less than the total bit vector size, which in our calculations was set to 8192 \newchanges{(SI Section~\ref{SI_subsec:substruct_to_struct_pubchem_data_curation})}. The transformer takes in the Morgan fingerprint bit vector through the encoder and then generates a multinomial distribution over the vocabulary of SMILES tokens, where the SMILES tokens are generated from SMILES strings using a regular expression\cite{schwaller_found_2018}. Sampling over the multinomial then produces the predicted SMILES strings.

For the spectrum-to-structure and spectrum-to-substructure multitask model (Figure~\ref{SI_fig:overview} (right)), the \textsuperscript{1}H NMR is represented as a vector of normalized intensities (a vector of size 28000 that evenly spans the NMR shift range from -2 to 12 ppm) which is processed through a two layer convolutional neural network embedding before being concatenated to an embedded version of the \textsuperscript{13}C NMR. 

\changed{The \textsuperscript{13}C NMR shifts are first rescaled using a minimum shift $\delta_{min}$ and maximum shift $\delta_{max}$ according to the following equation:}
\begin{equation}
    \textcolor{black}{\hat{\delta} = \frac{\delta - \delta_{min}}{\delta_{max} - \delta_{min}}}
\end{equation}
\changed{where in our work we use $\delta_{min}=-250$ ppm and $\delta_{max}=350$ ppm for maximal coverage of the \textsuperscript{13}C shift range. Each shift is then embedded into a vector using a linear transformation. To ensure that the model does not implicitly obtain information about carbons that would not be distinguishable due to symmetry, the shifts are de-duplicated to remove any identical shifts.} 

We process the two spectra modalities in different ways because of the different information that they encode: \textsuperscript{1}H NMR spectra contain finer splittings and peak shapes which encode important structural information about the molecule whereas the \textsuperscript{13}C NMR spectra are typically decoupled, making their peak shapes and intensities uninformative. Thus, we use a finer encoding for the \textsuperscript{1}H NMR and \changed{only embed the chemical shifts for the \textsuperscript{13}C NMR}. The model can optionally use either \textsuperscript{1}H NMR or \textsuperscript{13}C NMR only, and which modalities are used can be toggled by the arguments \texttt{use\_hnmr} and \texttt{use\_cnmr}, respectively, in the code provided with this work.

This joint representation is used as the input to both an encoder-decoder transformer model which generates the structure prediction as SMILES strings and the substructure prediction as a vector of probabilities, where each predicted probability represents how likely it is for that specific substructure to be present (see SI Section~\ref{SI_subsec:spec_to_struct_substruct_set}). To obtain the substructure prediction from the encoder network, we use sequence pooling\cite{hassani_escaping_2022} which performs the following mathematical operations in sequence: 
\begin{align}
    \mathbf{x}_L &= f(\mathbf{x}_0)\in\mathbb{R}^{N\times T \times E}\\
    \mathbf{x}'_L &= \text{softmax}(g(\mathbf{x}_L)^T)\in \mathbb{R}^{N \times 1 \times T}\\
    \mathbf{z} &= \text{squeeze}(\mathbf{x}'_L\mathbf{x}_L)\in \mathbb{R}^{N\times E}
\end{align}
where $N$ is the batch size, $T$ is the sequence length, $E$ is the embedding dimension, and $g(\cdot)\in \mathbb{R}^{E\times 1}$ is a learnable linear transformation. This can be understood as attending across the sequence dimension of the data after processing it through the encoder and assigning importance weights to each element in the sequence before aggregation. 

\changed{For initializing the multitask model from the pretrained substructure-to-structure transformer, we only transfer weights from the pretrained transformer into the structure prediction branch of the model (i.e., the encoder-decoder transformer) without transferring any of the pretrained weights into the substructure prediction branch of the multitask model (i.e. the encoder-only transformer). We found in experiments using a previous smaller version of the model with the dataset from our previous work\cite{hu_accurate_2024} that transferring the weights from the pretrained substructure-to-structure transformer into the substructure elucidation encoder portion of the multitask model had a negligible impact on both the structure elucidation accuracy and the substructure F1 score as shown in Table~\ref{SI_table:pretrained_encoder_results}.}

\begin{table}[!ht]
    \centering
    \caption{\changed{Model performance with and without using pretrained encoder weights. Model trained using \textsuperscript{1}H and \textsuperscript{13}C NMR as inputs.}}
    \label{SI_table:pretrained_encoder_results}
    \resizebox{\columnwidth}{!}{\begin{tabular}{ccc}
    \toprule
    \textbf{With pretrained weights} & \textbf{Top-15 structure accuracy (\%)} &\textbf{Substructure F1 score} \\ \midrule
    No & \numericalresult{70.3\%} & \numericalresult{0.87}\\
    Yes & \numericalresult{70.7\%} & \numericalresult{0.87}\\ \bottomrule
    
    \end{tabular}}
\end{table}

The specific hyperparameters for each portion of the model are compiled in Table~\ref{SI_table:substruct_to_struct_transformer_architecture} for the substructure-to-structure transformer model, Table~\ref{SI_table:multitask_encoder_architecture} for the spectrum-to-substructure encoder model, and Table~\ref{SI_table:multitask_conv_embed_architecture} for the convolutional embedding model for the spectra. We implement, train, and evaluate all of our models in PyTorch\cite{paszke_pytorch_2019} and PyTorch lightning. The original code is provided in the \changed{GitHub repository\cite{hu_nmrelucidator_nodate}}.

\begin{table}[!ht]
    \centering
    \caption{Substructure-to-structure architectural parameters, descriptions, and their values.}
    \label{SI_table:substruct_to_struct_transformer_architecture}
    \resizebox{\columnwidth}{!}{\begin{tabular}{lll}
        \toprule
        \textbf{Parameter} & \textbf{Description} & \textbf{Value} \\ \midrule
        \ttt{d\_model} & Embedding dimension of the model & 1024\\ 
        \ttt{d\_out} & Output dimension of the model & 2826\\
        \ttt{dim\_feedforward} & Hidden layer dimension of the feed forward neural network within the transformer blocks & 2048 \\
        \ttt{source\_size} & Total number of possible token values for embedding substructure sequences & 8193 \\
        \ttt{src\_pad\_token} & The index used for padding source sequences to the same length & 0\\
        \ttt{target\_size} & Total number of possible token values for embedding SMILES token sequences & 204\\
        \ttt{tgt\_pad\_token} & The index used for padding target sequences to the same length & 201 \\ 
        \ttt{num\_encoder\_layers} & The number of encoder layers & 8\\
        \ttt{num\_decoder\_layers} & The number of decoder layers & 8\\
        \ttt{nhead} & The number of heads used in multihead attention & 8\\
        \ttt{activation} & The activation function used for intermediate encoder/decoder layers & \ttt{relu}\\
        \ttt{dropout}\cite{hinton_improving_2012} & The probability for a particular element of an input tensor to be randomly set to 0 & 0.1\\
        \ttt{layer\_norm\_eps} & The constant used for numerical stability in layer normalization\cite{ba_layer_2016} & 1E-5\\ \midrule
    \end{tabular}}
\end{table}

\begin{table}[!ht]
    \centering
    \caption{Spectrum-to-substructure multitask model parameters, descriptions, and their values.}
    \label{SI_table:multitask_encoder_architecture}
    \resizebox{\columnwidth}{!}{\begin{tabular}{lll}
        \toprule
        \textbf{Parameter} & \textbf{Description} & \textbf{Value} \\ \midrule
        \ttt{d\_model} & Embedding dimension of the model & 1024\\ 
        \ttt{dim\_feedforward} & Hidden layer dimension of the feed forward neural network within the transformer blocks & 2048 \\
        \ttt{num\_encoder\_layers} & The number of encoder layers & 4\\
        \ttt{nhead} & The number of heads used in multihead attention & 4\\
        \ttt{activation} & The activation function used for intermediate encoder/decoder layers & \ttt{relu}\\
        \ttt{dropout}\cite{hinton_improving_2012} & The probability for a particular element of an input tensor to be randomly set to 0 & 0.1\\
        \ttt{layer\_norm\_eps} & The constant used for numerical stability in layer normalization\cite{ba_layer_2016} & 1E-5\\ \midrule
    \end{tabular}}
\end{table}

\begin{table}[!ht]
    \centering
    \caption{Spectrum convolutional embedding model parameters, descriptions, and their values.}
    \label{SI_table:multitask_conv_embed_architecture}
    \resizebox{\columnwidth}{!}{\begin{tabular}{lll}
        \toprule
        \textbf{Parameter} & \textbf{Description} & \textbf{Value} \\ \midrule
        \ttt{d\_model} & Output dimension of the spectrum embeddings & 1024\\ 
        \ttt{n\_hnmr\_features} & Number of features to embed for the \textsuperscript{1}H NMR spectrum & 28000\\
        \ttt{n\_cnmr\_features} & Number of features to embed for the \textsuperscript{13}C NMR spectrum & 80\\
        \ttt{use\_hnmr} & Use the \textsuperscript{1}H NMR features as input & \ttt{True}\\
        \ttt{use\_cnmr} & Use the \textsuperscript{13}C NMR features as input & \ttt{True}\\
        \ttt{h\_dropout\_rate} & Dropout probability for \textsuperscript{1}H NMR logits & 0.0 \\
        \ttt{c\_dropout\_rate} & Dropout probability for \textsuperscript{13}C NMR logits & 0.0 \\
        \ttt{pool\_variation} & The pooling downsampling operation to use between the convolutional layers & \ttt{max} \\
        \ttt{kernel\_size\_1} & The convolutional kernel size for the first convolutional layer & 5 \\
        \ttt{out\_channels\_1} & The output number of filters after the first convolutional layer & 64 \\ 
        \ttt{pool\_size\_1} & The kernel size for the first pooled downsampling & 12 \\
        \ttt{kernel\_size\_2} & The convolutional kernel size for the second convolutional layer & 9 \\ 
        \ttt{out\_channels\_2} & The output number of filters after the second convolutional layer & 128 \\ 
        \ttt{pool\_size\_2} & The kernel size for the second pooled downsampling & 20 \\
        \ttt{add\_pos\_encode} & Whether positional encoding information is added & \ttt{True} \\
        \changed{\ttt{cnmr\_max\_shift}} & \changed{The maximum shift value for rescaling, in ppm} & \changed{350} \\
        \changed{\ttt{cnmr\_min\_shift}} & \changed{The minimum shift value for rescaling, in ppm} & \changed{-250}\\
        \midrule
    \end{tabular}}
\end{table}

\section{Details on the substructure-to-structure transformer}\label{SI_sec:substruct_to_struct}
Here we provide additional details on how the substructure-to-structure transformer was trained and evaluated, including the data curation, Morgan fingerprint representation, sampling protocol, and additional analyses.

\subsection{Morgan fingerprint representation scans}\label{SI_subsec:substruct_to_struct_mfp_scan}
When constructing a Morgan fingerprint, there are two parameters that must be chosen: the total fingerprint vector size and the radius of the atomic environments to consider. To decide on the final Morgan fingerprint representation for training the model, we used the transformer size and architecture and the set of 3M molecules from our previous work\cite{hu_accurate_2024} and trained variations of the substructure-to-structure transformer using fingerprints computed using different radii and bit vector sizes. The results are summarized in Table~\ref{SI_table:substruct_to_struct_fingerprint_scan}, where we found that a fingerprint radius of 2 and a bit vector size of 8192 were the best choice to achieve the greatest amount of accuracy without introducing excessively many tokens into the transformer's substructure vocabulary. 

\begin{table}[!ht]
    \centering
    \caption{Substructure-to-structure transformer accuracy as a function of both the fingerprint size and fingerprint radius used to represent the substructures. The best result is bolded.}
    \label{SI_table:substruct_to_struct_fingerprint_scan}
    \sisetup{round-mode=places}
    \small
    \resizebox{0.8\columnwidth}{!}{\begin{tabular}{cS[round-precision=2]S[round-precision=2]S[round-precision=2]S[round-precision=2]}
        \toprule
        \textbf{Fingerprint size} & \textbf{Radius 2} & \textbf{Radius 3} & \textbf{Radius 4} & \textbf{Radius 5} \\ \midrule
        1024 & 93.52\% & 91.78\% & 91.06\% & 90.83\% \\
        2048 & 95.55\% & 94.60\% & 94.14\% & 94.27\% \\
        4096 & 96.77\% & 96.28\% & 96.01\% & 95.99\% \\
        8192 & \bf{97.44\%} & 97.27\% & 97.18\% & 97.26\% \\ \midrule
    \end{tabular}}
\end{table}

\subsection{Dataset curation from PubChem}\label{SI_subsec:substruct_to_struct_pubchem_data_curation}
\begin{figure}[!h]
    \centering
    \includegraphics[width=\textwidth]{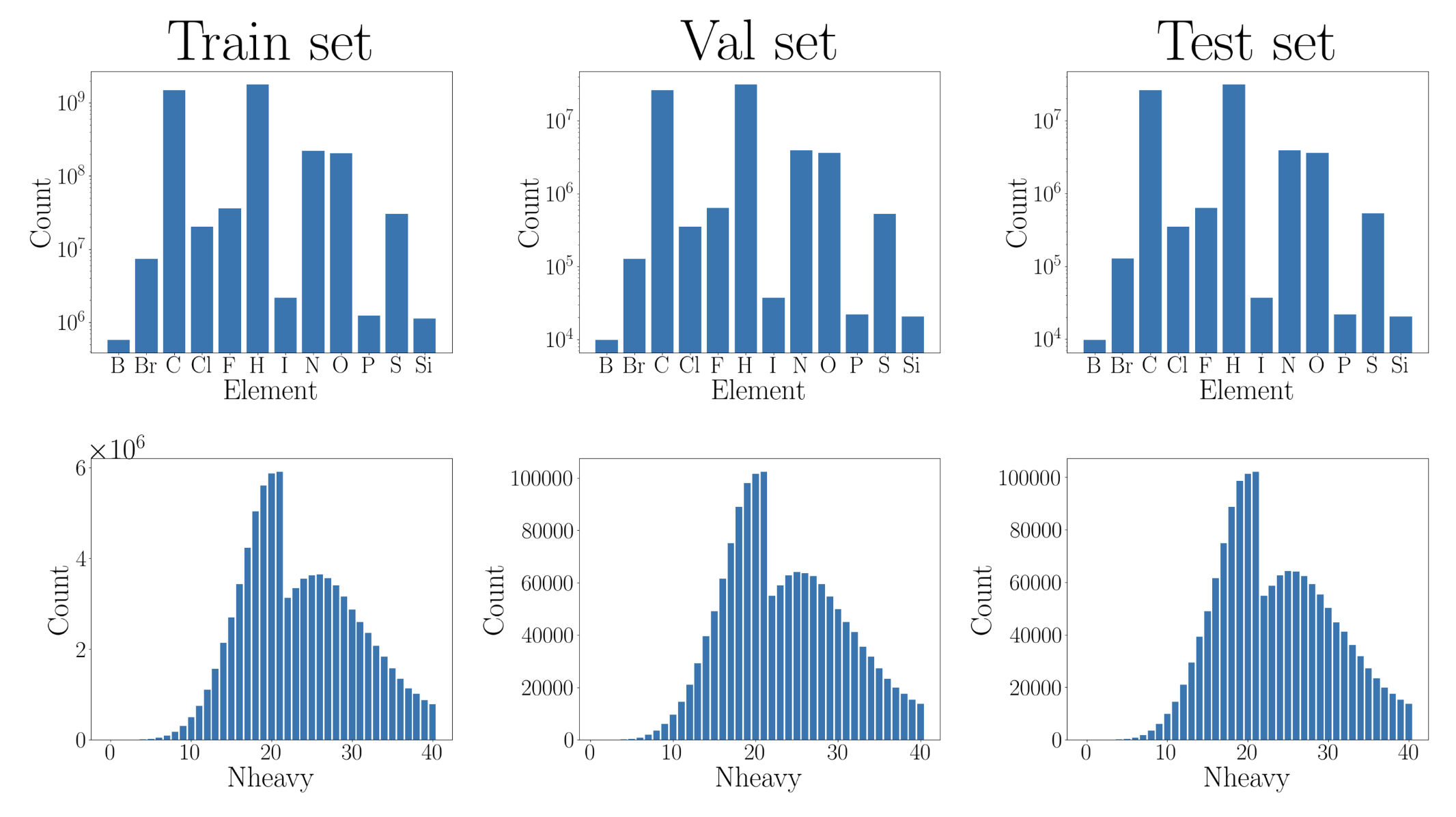}
    \caption{(Top) The frequencies of different elements in the training, validation, and test sets of the 88M PubChem dataset and (Bottom) the distribution of molecular sizes in terms of number of heavy atoms.}
    \label{SI_fig:pubchem_size_elem_dist}
\end{figure}

To assemble the dataset for training the substructure-to-structure transformer, we first downloaded all the CID SMILES strings available in PubChem\cite{kim_pubchem_2025} which gave a total of $\sim$171M SMILES strings as of February 2025. We then canonicalized these SMILES strings, removed any invalid entries (i.e., not parsable by RDKit\cite{noauthor_rdkit_nodate} into a valid molecule), and then further filtered these strings down by chemical composition, removing molecules which contained elements other than C, N, O, H, B, P, S, Si, F, Br, Cl, and I. This left a set of 103M SMILES strings. After removing \changed{stereochemical configuration} from the SMILES strings and further removing duplicates due to stereoisomers, this yielded the final set of 88M SMILES strings used for training the substructure-to-structure transformer. The elemental composition and distribution of molecule sizes are shown in Figure~\ref{SI_fig:pubchem_size_elem_dist}. We attribute the skew of the molecule sizes towards systems with $\leq$21 heavy atoms to Lipinski's rule of five\cite{lipinski_lead-_2004}, a set of constraints which encourage oral bioactivity for drug-like molecules. The data was then randomly partitioned into a train, validation, and test set, with \changed{96.6\%} for training, 1.7\% for validation, and 1.7\% for testing. Specifically, 85269445 SMILES strings were used for training, 1505309 for validation, and 1505601 for testing. 

\begin{figure}[!h]
    \centering
    \includegraphics[width=\textwidth]{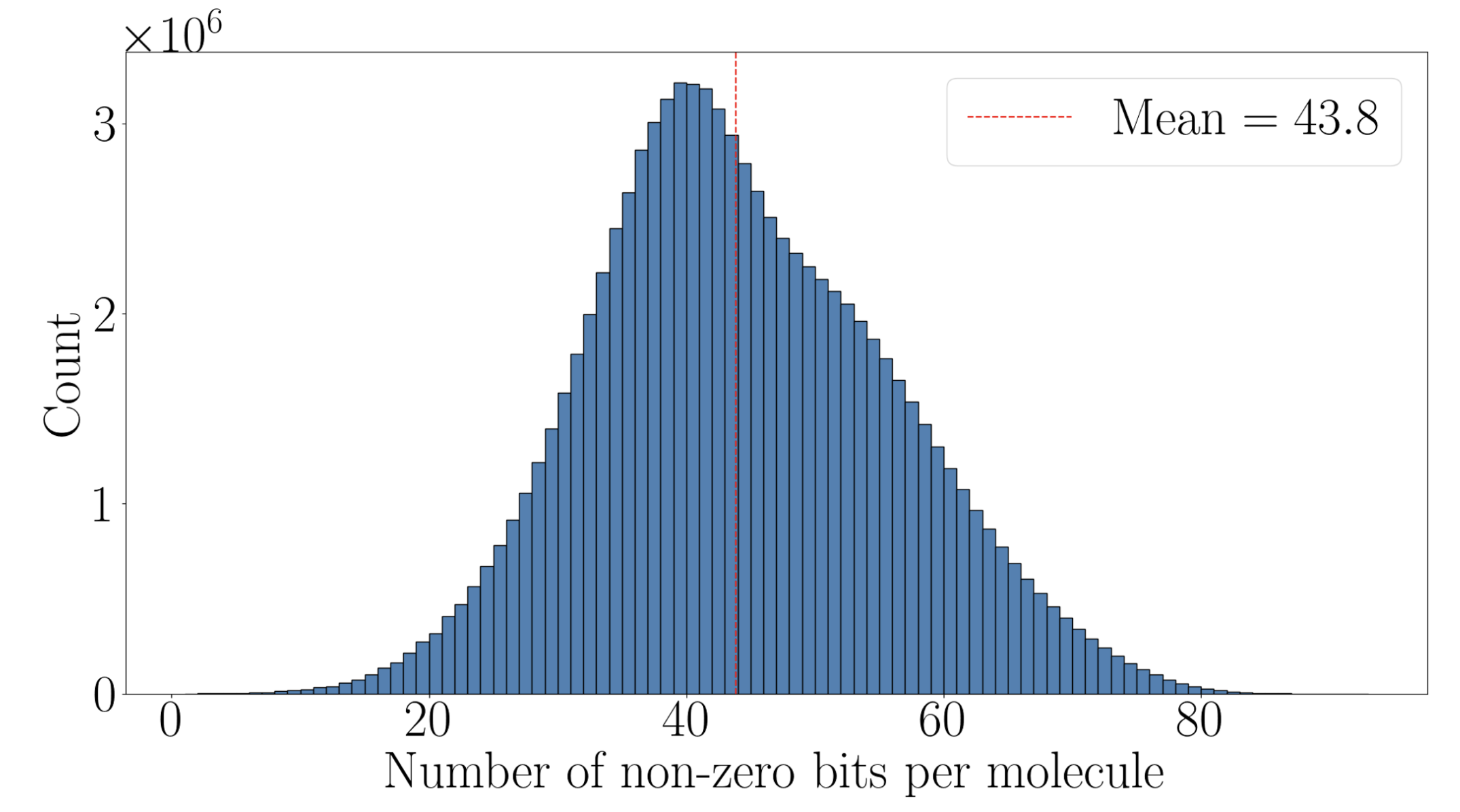}
    \caption{\newchanges{Distribution of the number of on bits for all 88M molecules in the PubChem dataset. On average, each molecule only has 43.8 on bits.}}
    \label{SI_fig:mfp_onbit_histogram}
\end{figure}

\newchanges{For all of the molecules in our dataset, we compute their Morgan fingerprints with a radius of 2 and a bit vector size of 8192. While the input token vocabulary size of the substructure-to-structure transformer is determined by the vector size used (the vocabulary is always equal to $n + 1$ for $n$ many bits and a padding token), because we only extract the on bits and embed those directly, the actual number of tokens used for each molecule is significantly less than 8192. On average, across all the 88M SMILES strings in the dataset, the average number of on bits per molecule is \numericalresult{43.8} (Figure~\ref{SI_fig:mfp_onbit_histogram}), with the maximum reaching only 93 on bits. This ensures that despite having the capacity to represent each of the 8192 bits in the Morgan fingerprint, each pass through the transformer does not need to consume 8192 embedding vectors, which would lead to significantly slower training.}

\subsection{Sampling protocol}\label{SI_subsec:substruct_to_struct_sampling}
To sample the SMILES strings of molecules from our trained substructure-to-structure transformer, we sampled from the transformer using top-$k$ random sampling\cite{fan_hierarchical_2018} with $k=5$. Thus at each step of sampling when generating a SMILES string, we do not sample from the full multinomial distribution over the tokens but instead sample from a smaller normalized distribution comprised of the tokens with the 5 highest probabilities. We used the sampling temperature of $T=1.0$ in all of our experiments. For each input Morgan fingerprint, we sampled 15 complete SMILES strings, where the model naturally terminates its generations by the prediction of a stop token. Using these 15, we then canonicalized the SMILES strings before computing the structure accuracy. As discussed in the main text, we only considered a molecule correctly predicted if its exact canonical SMILES string appears within the set of 15 predictions. 

\subsection{Optimization protocol}\label{SI_subsec:substruct_to_struct_training}
To optimize the substructure-to-structure transformer, we used the AdamW optimizer with a constant learning rate of $2.0\times10^{-5}$, $\beta=(0.9, 0.98)$, $\epsilon=1.0\times10^{-9}$, and a weight decay of 0.01. The model was trained on 4 NVIDIA H100 GPUs in 16-mixed precision for a total of 2000000 steps with a batch size of 64, with the validation loss measured every 2000 steps. Gradient clipping was enabled on the gradient norm, with a gradient clip value of 1.0. The checkpoint with the best validation loss was used for all the evaluations of the substructure-to-structure task, as well as the starting point for initializing the multitask framework.

\newpage
\subsection{Additional figures and analysis}\label{SI_subsec:substruct_to_struct_addn_figs}

\begin{figure}[!h]
    \centering
    \includegraphics[width=\textwidth]{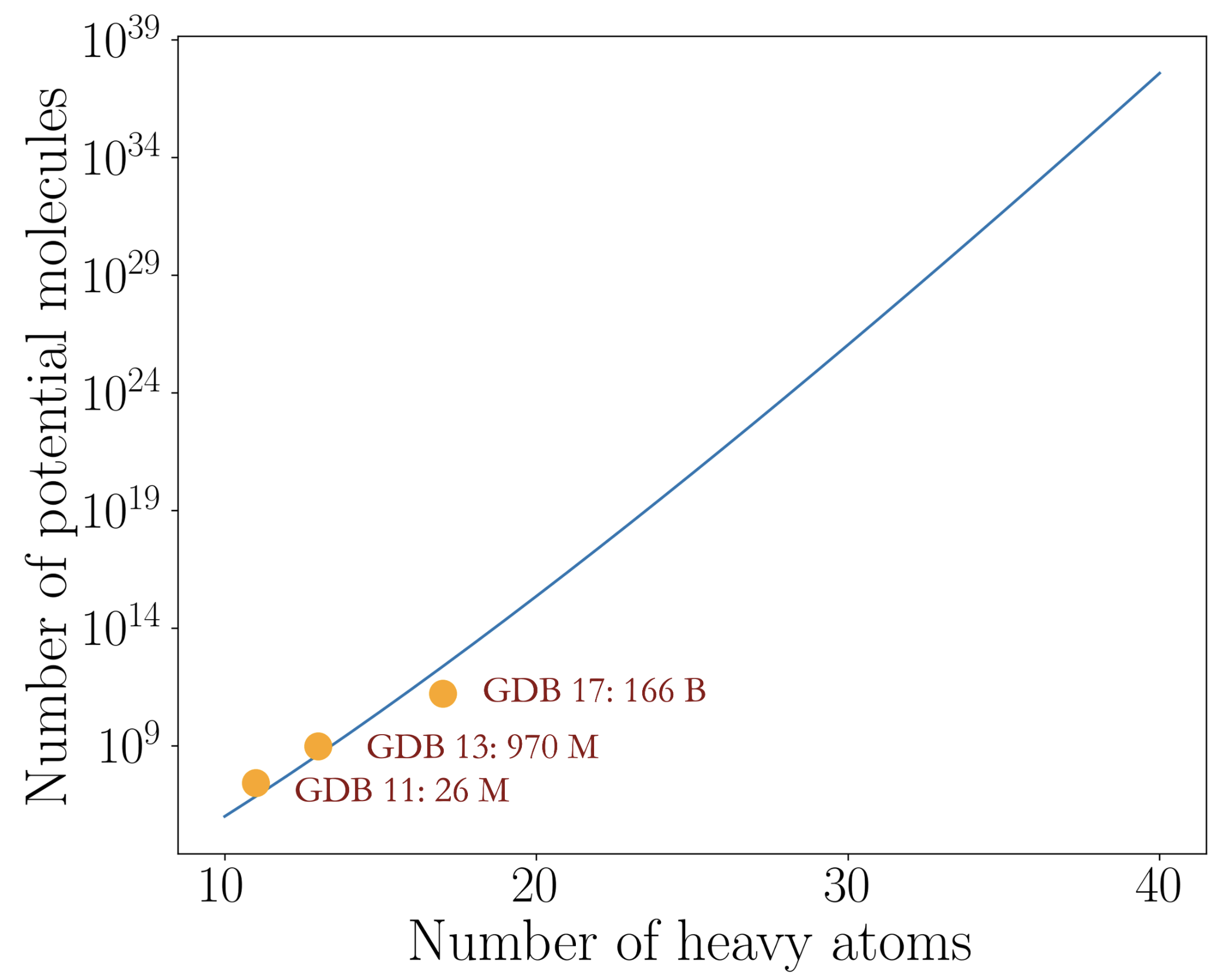}
    \caption{Extrapolation of the size of chemical space as a function of the number of heavy atoms based on the empirical rule presented in Polishcuk et al.\cite{polishchuk_estimation_2013}. The sizes of the GDB 11\cite{fink_virtual_2007}, GDB 13\cite{blum_970_2009}, and GDB 17\cite{ruddigkeit_enumeration_2012} datasets are plotted for comparison.}
    \label{SI_fig:chemical_space_extrapolation}
\end{figure}

\begin{table}[!h]
    \centering
    \caption{Statistics of the distributions of the MTS values for incorrect predictions for different ranges of heavy atoms. Increasing size leads to both higher average MTS and a greater percentage above a 0.5 similarity threshold.}
    \label{SI_table:substruct_to_struct_stats_tanimoto_dist_nheavies}
    \sisetup{round-mode=places}
    \large
    \resizebox{\columnwidth}{!}{\begin{tabular}{cS[round-precision=2]c}
        \toprule
        \textbf{Heavy atom range} & \textbf{Avg. MTS} & \textbf{\% of MTS values similarities above 0.5} \\ \midrule
        $0 \leq n < 10$ & \numericalresult{0.66} & \numericalresult{79.3}\\ 
        $10 \leq n < 20$ & \numericalresult{0.77} & \numericalresult{95.6} \\ 
        $20 \leq n < 30$ & \numericalresult{0.83} & \numericalresult{98.7} \\ 
        $30 \leq n \leq 40$ & \numericalresult{0.86} & \numericalresult{99.2} \\ \midrule
    \end{tabular}}
\end{table}

\begin{figure}[!h]
    \centering
    \includegraphics[width=0.8\textwidth]{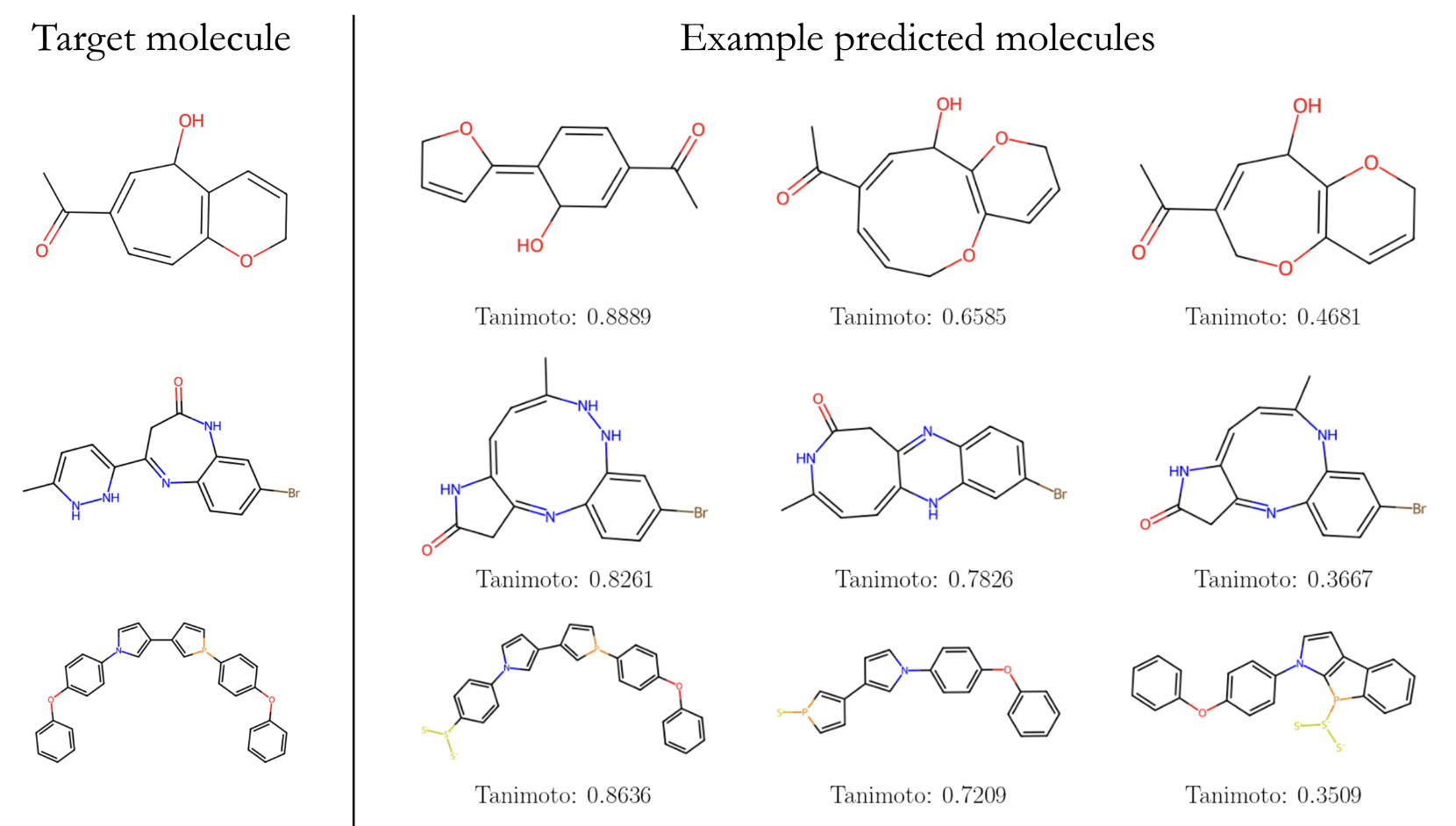}
    \caption{Examples of target molecules and incorrect predictions with different Tanimoto similarities to the target. The incorrect predictions were de-duplicated and three molecules were chosen specifically in the ranges of (0.8 - 0.9), (0.5 - 0.8), and (0.0 - 0.5).}
    \label{SI_fig:substruct_to_struct_targs_and_incorrect}
\end{figure}

\section{Details on the spectrum-to-structure multitask framework}\label{SI_sec:spec_to_struct}
Here we provide additional details on the spectrum-to-structure and spectrum-to-substructure multitask framework, including the data curation strategy for generating the dataset used for training and the determination of the substructure set.

\subsection{Dataset curation and subsampling approach}\label{SI_subsec:spec_to_struct_curation}
To create a diverse set of molecules to compute the NMR spectra for starting with the full set of 171M PubChem SMILES strings (SI section~\ref{SI_subsec:substruct_to_struct_pubchem_data_curation}), we first subsampled a set of 103M SMILES strings, which included stereoisomers, down to a set of 5M SMILES strings without considering \changed{stereochemical configuration} using Algorithm~\ref{SI_algo:pubchem_sample_to_5M}. In this algorithm, we build up the dataset by randomly sampling a molecule from the 103M SMILES strings and then deciding whether to add it to the current dataset by comparing its Morgan fingerprint to the population distribution of fingerprints bits currently in the dataset. We use this comparison to bias the random sampling towards molecules whose substructures are under-represented in the current set and thereby move away from sampling molecules that are similar.  This iterative procedure was repeated until a dataset of the required size of 5M was obtained. The filtering function, $G(\cdot)$, in Algorithm~\ref{SI_algo:pubchem_sample_to_5M} filters out molecules that contain radicals, ions, or isotopes, thereby ensuring that the dataset is comprised of closed-shell, neutrally charged organic molecules. The fingerprints used for this procedure were all of radius 2, with a bit vector size of 8192 (SI section~\ref{SI_subsec:substruct_to_struct_mfp_scan}).

After this initial algorithm to reduce the PubChem set from 103M to 5M, a second step of filtering was applied to reduce the final set to 2M molecules. Here, we employed the recently developed BitBirch program\cite{perez_bitbirch_2025} which enables rapid clustering of large molecular libraries based on efficient estimations of molecular similarity. This removes the need to compute all $\mathcal{O}(N^2)$ comparisons across the set of 5M molecules, which is prohibitively expensive, or any reliance on approximate estimations of internal diversity. Our approach for selecting the final set of 2M molecules is outlined in Algorithm~\ref{SI_algo:bitbirch_cluster_sample}, where we performed 5 rounds of the clustering and sampling procedure with BitBirch. The motivation for using this clustering approach is to restrict the expensive calculation of pairwise similarity ($\mathcal{O}(N^2)$ cost) to only smaller clusters of similar molecules, enabling better identification of highly similar SMILES strings that need to be removed. To further accelerate this sampling procedure, we opted for a smaller but still expressive fingerprint size of 2048, with a radius of 2. This produced a set of SMILES strings slightly larger than 2M. 

To obtain the final 2M set, we randomly sampled 1600000 of training, 200000 validation, and 200000 test set SMILES from the larger training, validation, and test sets obtained after the clustering and filtering procedure, ensuring that each randomly sampled molecule contained at least one non-exchangeable proton. We performed an additional step of repartitioning this set of 2M SMILES based on the train-val-test splitting of the 143K set of spectra from our previous work\cite{hu_accurate_2024} to ensure that there is no dataset leakage when testing our new model on the old test set, resulting in the numbers of training, validation, and test set molecules that deviate slightly from the exact 80-10-10 split. \newchanges{We forward simulate the \textsuperscript{1}H NMR spectra and \textsuperscript{13}C NMR chemical shifts for all of the molecules using the v2024.2 batch NMR predictors from ACD/Labs. The \textsuperscript{1}H NMR spectra were predicted using a field strength of 500.12 MHz with a total grid size of 32768 points equally spaced between -2 to 12 ppm with exchangeable protons excluded. A neural network-based approach was used for the prediction of the chemical shifts, and HOSE\cite{bremser_hose_1978} codes were used for the coupling constants, going up to \textsuperscript{4}J in couplings with no specific solvent (i.e., using the average of all solvents that were in the training set). The \textsuperscript{13}C NMR shift predictions were performed using a neural network-based approach at 125.03 MHz from -20 to 230 ppm, again with no specific solvent. We did not use solvent-specific predictions as these would affect more the exchangeable protons in the structure, which are excluded. The resulting effect of using the average of all the solvents in the training set is minimal for \textsuperscript{13}C NMR prediction and for \textsuperscript{1}H NMR prediction, the advantage of having a broader set of structures for training the neural network predictor overcomes the smaller disadvantage of having slightly higher deviations in values for protons next to atoms with exchangeable protons.}

\begin{figure}[!h]
    \centering
    \includegraphics[width=\textwidth]{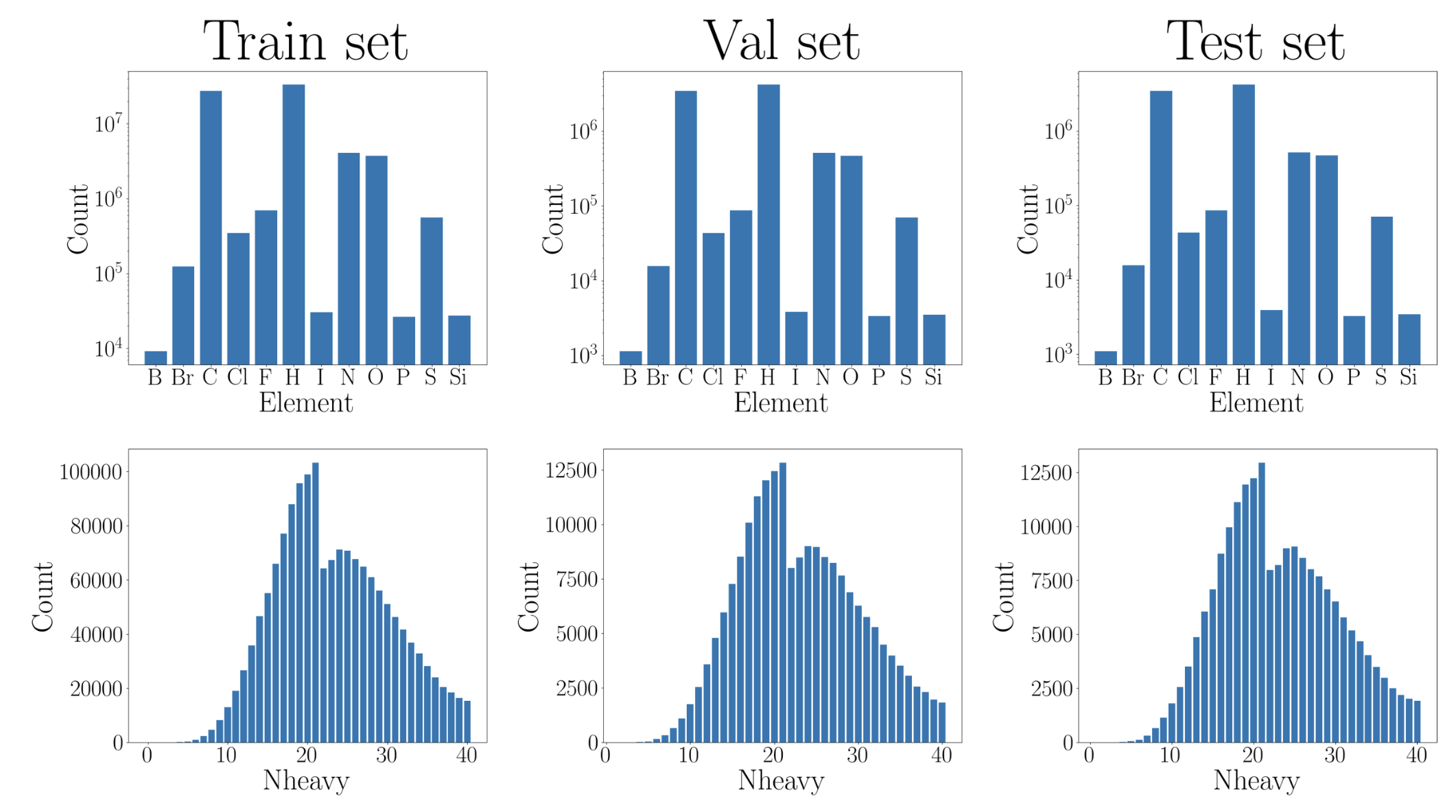}
    \caption{(Top) The frequencies of different elements in the training, validation, and test sets of the 2M subsampled dataset and (Bottom) the distribution of molecular sizes in terms of number of heavy atoms.}
    \label{SI_fig:spec_to_struct_size_elem_dist}
\end{figure}

After forward simulating and collating all of the spectra, the final dataset consisted of 1595793 SMILES and spectra for training, 202151 for validation, and 201953 for testing. This dataset was then used for training and evaluation of the spectrum-to-structure and spectrum-to-substructure multitask framework. The elemental composition and distribution of molecule sizes are shown in Figure~\ref{SI_fig:spec_to_struct_size_elem_dist}.

\begin{fullwidth}[width=\linewidth+2cm,leftmargin=-1cm,rightmargin=-1cm]
\begin{algorithm}[H]
\caption{Sampling 5M SMILES from PubChem}\label{SI_algo:pubchem_sample_to_5M}
\begin{algorithmic}
\Require Dataset $\mathcal{D}$, Number to sample $n$, Fingerprint generator $F(\cdot)$ that generates vectors of size $M$, Filtering function $G(\cdot)$, \changed{Stereochemical configuration} removal function $H(\cdot)$, Number of maximum iterations $N$
\Ensure New dataset $\mathcal{D}'$ with $n$ molecules, Array of sizes $P$ for all molecules in $\mathcal{D}'$
\State $\mathcal{D}' \gets [ \ ]$
\Comment{Initialize with empty list}
\State $n_{iter} \gets 0$
\State $counts \gets null$ \Comment{Cumulative bit count fingerprint}
\State $S \gets \{ \ \}$ \Comment{Empty set for tracking seen SMILES}
\State $P \gets [ \ ]$

\While{($n_{iter} < N$) and (\text{length}($\mathcal{D}') < n$)}
\State{$s \sim \mathcal{D}$} \Comment{Sample a SMILES uniformly from dataset}
\State{$s \gets H(s)$}
\If {
($s$ is not null) and ($s \notin S$) and ($G(s) = \text{True}$)
}
    \State{$mol \gets \text{SmilesToMol}(s)$}
    \State{$fp \gets F(mol)$} \Comment{$\in \{0, 1\}^M$}
    \If{($counts$ is null)}
        \State{$counts \gets fp$}
    \EndIf
    \State{$norm \gets \frac{counts}{\sum_i counts_i}$} \Comment{$\in (0,1)^{M}$}
    \State{$score \gets fp \cdot norm$} \Comment{$\in (0, 1)$}
    \State{$u \sim U(0, 1)$}
    \If {$u < (1 - score)$}
        \State{$\mathcal{D'} \gets \mathcal{D}' \text{ append } s$}
        \State{$P \gets P \text{ append } mol.\text{GetNumHeavyAtoms()}$}
        \State{$S \gets S \ \bigcup \ \{s\}$}
        \State{$counts \gets counts + fp$} \Comment{$\in \mathbb{N}^M$}
    \EndIf
\EndIf
\State{$n_{iter} \gets n_{iter} + 1$}
\EndWhile
\State \Return $\mathcal{D}'$, $P$
\end{algorithmic}
\end{algorithm}
\end{fullwidth}

\begin{fullwidth}[width=\linewidth+2cm,leftmargin=-1cm,rightmargin=-1cm]
\begin{algorithm}[H]
\caption{Sampling 2M SMILES with BitBirch clustering}
\label{SI_algo:bitbirch_cluster_sample}
\begin{algorithmic}
\Require A total set of SMILES strings partitioned into a training, validation, and test set $\boldsymbol{S}=[S_{train}, S_{val}, S_{test}]$, maximum number of sweeps $N$, minimum number of train SMILES $n_{train}$, validation SMILES $n_{val}$, and test SMILES $n_{test}$, similarity threshold for pruning $t$. 
\Ensure Filtered set of SMILES $\boldsymbol{S'}=[S'_{train}, S'_{val}, S'_{test}]$
\State $n_{iter} \gets 0$
\State $\boldsymbol{S'} \gets \boldsymbol{S}$
\While{($n_{iter} < N$) and all($[|S_{train}|, |S_{val}|, |S_{test}|] \geq [n_{train}, n_{val}, n_{test}]$)}
    \State {$\boldsymbol{C} \gets \text{BitBirchCluster}(\boldsymbol{S'})$} \Comment{Cluster labels for all SMILES}
    \State {$\boldsymbol{D} \gets \text{AggregateClusters}(\boldsymbol{C})$} \Comment{Maps cluster indices to SMILES indices}
    \For{$D_i\in \boldsymbol{D}$}
        \If{$|D_i| > 1$} \Comment{Cluster contains more than 1 SMILES}
            \State {$\text{pairs}, \text{sims} \gets \text{AllPairwiseSimilarity}(D_i)$}
            \For{$p_k\in\text{pairs}$, $sim_k \in\text{sims}$}
                \If{$sim_k > t$}
                    \State{$i,j\gets p_k$}
                    \State{$s_i, s_j \gets \boldsymbol{S'}[i], \boldsymbol{S'}[j]$}
                    \State{$\boldsymbol{S'} \gets \text{SetRemove}(\boldsymbol{S'}, s_i, s_j)$} \Comment{Remove the $s_l$ belonging to largest of $S'_{train}, S'_{val}, S'_{test}$}
                \EndIf
            \EndFor
        \EndIf
    \EndFor 
\EndWhile
\State \Return $\boldsymbol{S'}$
\end{algorithmic}
\end{algorithm}
\end{fullwidth}

\subsection{Generation of a new substructure set}\label{SI_subsec:spec_to_struct_substruct_set}
In order to predict substructures from the multitask framework, the model must have a persistent representation of substructures that can be interpreted independently of the identity of a given molecule. Because of this, Morgan fingerprints are not a suitable representation for substructure prediction because individual bits of a Morgan fingerprint cannot be uniquely mapped back to a substructure without knowing the identity of the full molecule. This limitation arises because of how Morgan fingerprints are constructed\cite{rogers_extended-connectivity_2010}, where circular substructure environments are extracted, hashed, and then allocated in a fingerprint, a process that can lead to hash collisions between different substructures and thus bits corresponding to certain environments appearing in different locations in the bit vector for different molecules. Thus, while the model can predict Morgan fingerprints directly, this is not ideal as such an output would lack physical interpretability. 

As an alternative to create a set of substructures that is consistent across molecules, we used the Morgan algorithm to extract a standalone set of substructures from the 2M SMILES strings used to train the spectrum-to-structure multitask framework, where each substructure is an atomic environment of radius 1. We then removed any substructures which occurred fewer than 50 times across the entire dataset. \changed{Since substructures can occur multiple times within a single molecule, the threshold of 50 assumes an overly optimistic scenario that each occurrence is in a new molecule, which is not guaranteed to be true. This threshold therefore corresponds to a maximum fraction of occurrence across the 2M molecules of $50 / 2000000 = 2.5\times10^{-5}$, and applying this threshold left a set of 2826 substructures that the model predicts the probabilities for}. 

\changed{To ensure that our set of 2826 substructures is well-converged for our task, we repeated the substructure extraction and analysis using the 88M PubChem SMILES strings dataset, applying the same threshold (a maximum fraction of occurrence of $2.5\times 10^{-5}$). Upon removing charged subgroups which were excluded by design when curating our set of 2M strings, there are only 41 substructures in the set of 2484 substructures derived from the PubChem SMILES that were not already in the set derived from the 2M molecules. This is a small difference: 1.5\% of the total set of 2826 substructures. Of the remaining 41, only 13 have a maximum fraction of occurrence greater than $5.0\times10^{-5}$, corresponding to less than 0.5\% of the 2826 substructure set. This shows that even though the subsampled 2M molecules is a smaller dataset than the 88M PubChem SMILES string, the molecules are diverse enough such that substructures which are not exceedingly rare are well-represented in both sets with negligible differences. Furthermore, since the NMR spectra data only covers the 2M set of molecules, the substructure set should place a greater emphasis on the substructures that are well-represented in this subset. We therefore use our set of 2826 substructures derived from the 2M molecule set for all experiments.}

\begin{figure}[!h]
    \centering
    \includegraphics[width=\textwidth]{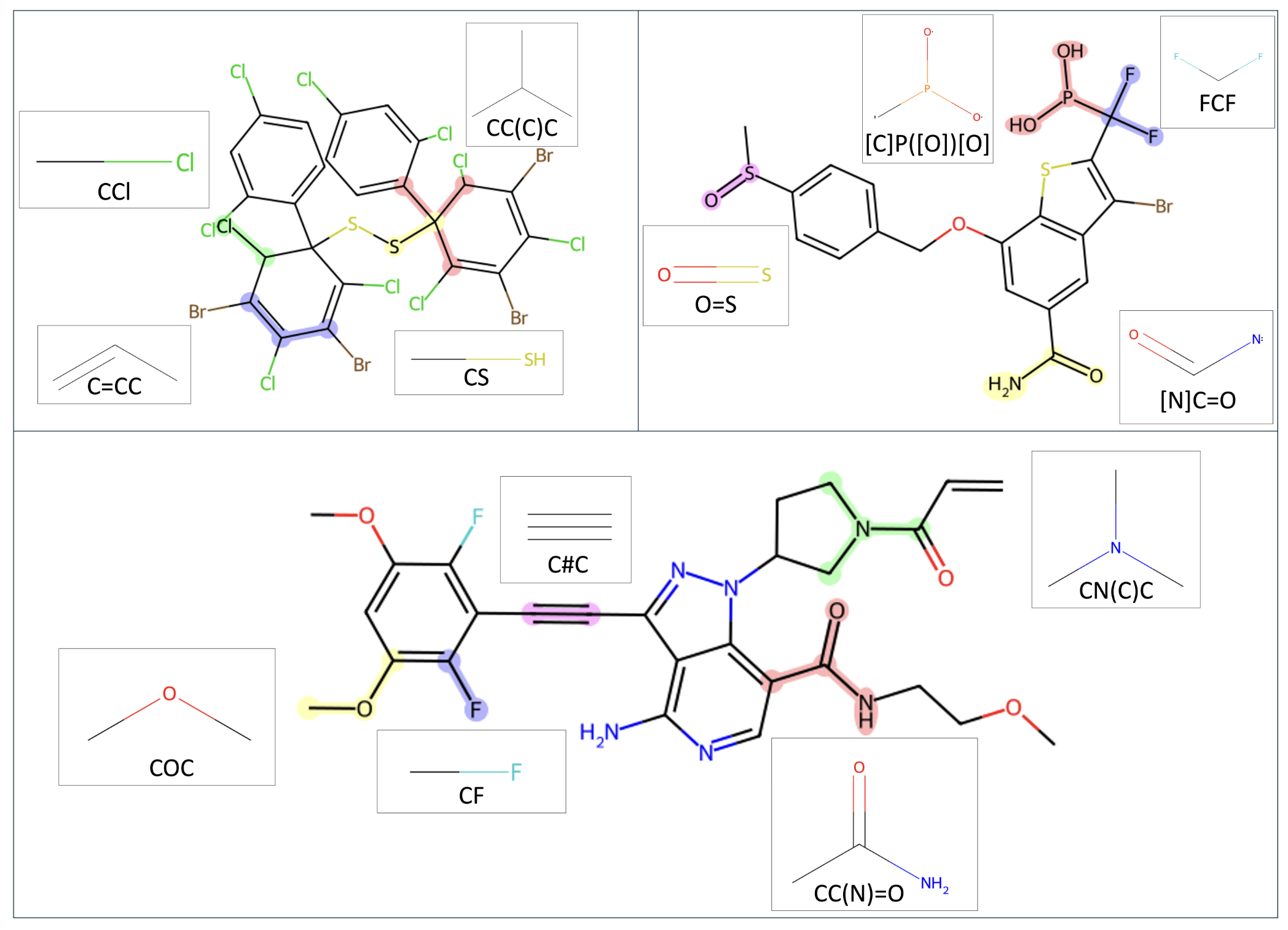}
    \caption{Examples of molecules with some substructures highlighted. The substructures, as well as their SMILES string representations, are provided in the boxes surrounding each molecule.}
    \label{SI_fig:spec_to_struct_highlighted_mols}
\end{figure}

Some examples of molecules with some of their substructures highlighted are shown in Figure~\ref{SI_fig:spec_to_struct_highlighted_mols}. The strings for all 2826 substructures, as well as the code for visualizing them, are provided in the \changed{Zenodo upload\cite{hu_datasets_2026}} with this work.

\begin{figure}[!h]
    \centering
    \includegraphics[width=0.8\textwidth]{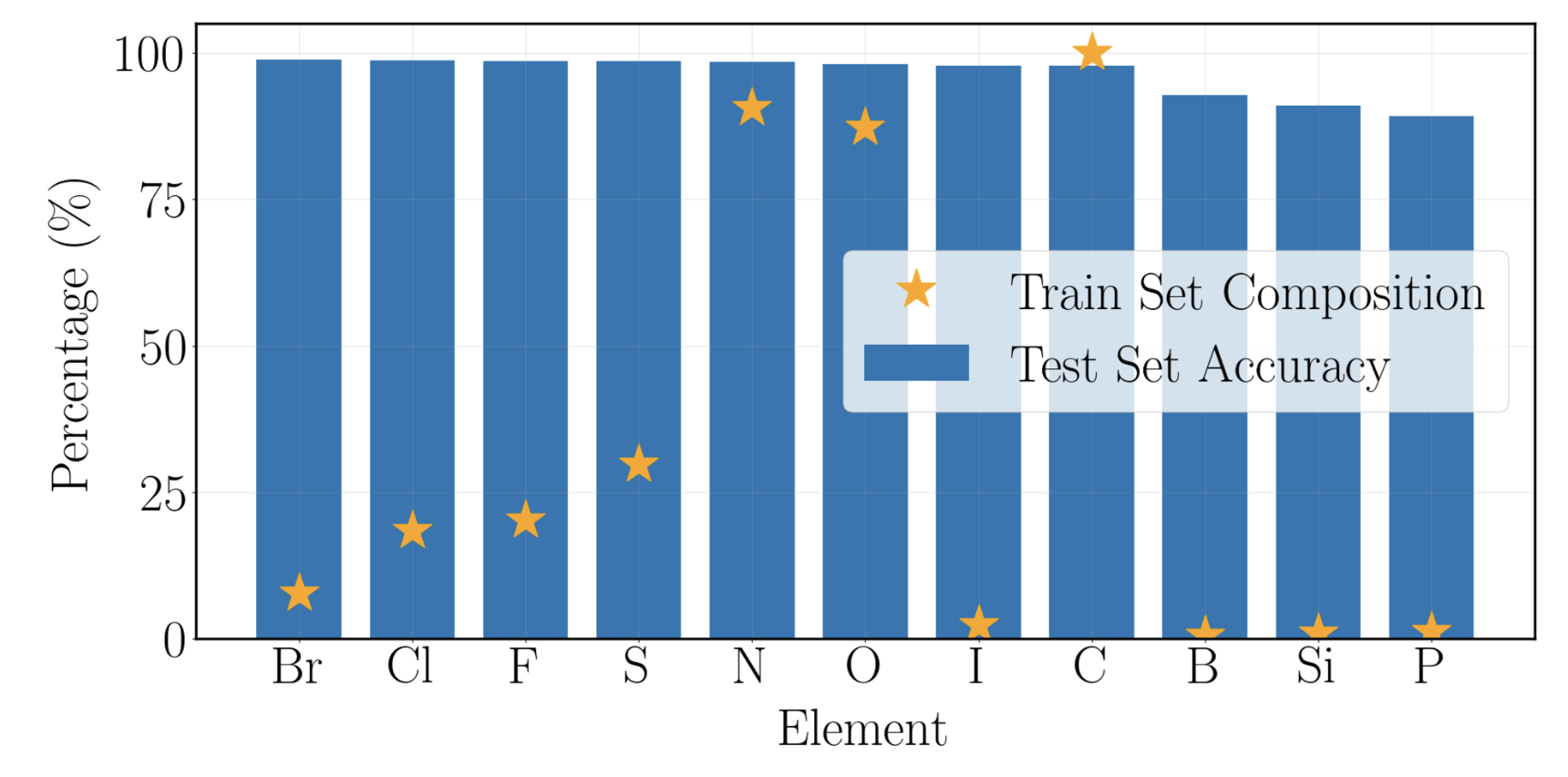}
    \caption{Test set accuracy of the substructure-to-structure transformer resolved by elemental composition. Notably, phosphorus (P) containing systems perform the worst out of the heavy atoms, indicating difficulties in generalizing to P-containing systems.}
    \label{SI_fig:substruct_to_struct_perf_by_elem}
\end{figure}

\subsection{Sampling protocol}\label{SI_subsec:spec_to_struct_sampling}
For generating SMILES strings from the trained multitask framework, we sample and compute accuracies using the same methods as in the substructure-to-structure case (SI Section~\ref{SI_subsec:substruct_to_struct_sampling}), with top-$k$ sampling with $k=5$ and a sampling temperature of $T = 1.0$ for \changed{N SMILES strings per NMR spectrum/spectra input depending on the metric being computed (e.g., Top-15 or Top-30)}. The substructure probabilities were predicted in one shot.

\subsection{Optimization protocol}\label{SI_subsec:spec_to_struct_training}
To optimize the spectrum-to-structure/spectrum-to-substructure multitask framework, we use the AdamW optimizer with a constant learning rate of $1.0\times10^{-5}$, $\beta=(0.9, 0.999)$, $\epsilon=1.0\times10^{-8}$, and a weight decay of 0.01. Each of the models presented in the main text were trained on 2 NVIDIA H100 GPUs with a batch size of 64 in single precision, with the lowest validation loss checkpoint used for evaluation.

The loss function optimized for the multitask framework was a linear combination of a binary cross entropy loss for the substructure prediction and a cross entropy loss for the SMILES string prediction:
\begin{equation}
    \mathcal{L}(\boldsymbol{\hat{y}}_{smi}, \boldsymbol{y}_{smi}, \boldsymbol{\hat{y}}_{sub}, \boldsymbol{y}_{sub}) = \alpha\mathcal{L}_{CE}(\boldsymbol{\hat{y}}_{smi}, \boldsymbol{y}_{smi})+\beta\mathcal{L}_{BCE}(\boldsymbol{\hat{y}}_{sub}, \boldsymbol{y}_{sub}) 
\end{equation}
where $\boldsymbol{\hat{y}}_{smi}$ and $\boldsymbol{y}_{smi}$ are the SMILES string \changed{prediction and target}, respectively, and $\boldsymbol{\hat{y}}_{sub}$ and $\boldsymbol{y}_{sub}$ are the substructure \changed{prediction and target}, respectively. $\alpha$ and $\beta$ are weighting factors which for joint optimization we set to be $\alpha=\beta=1$. 

\subsection{\changed{Top-N and Rank-1-of-N metrics}}\label{SI_subsec:expanded_acc_metrics}
\begin{figure}[!h]
    \centering
    \includegraphics[width=\textwidth]{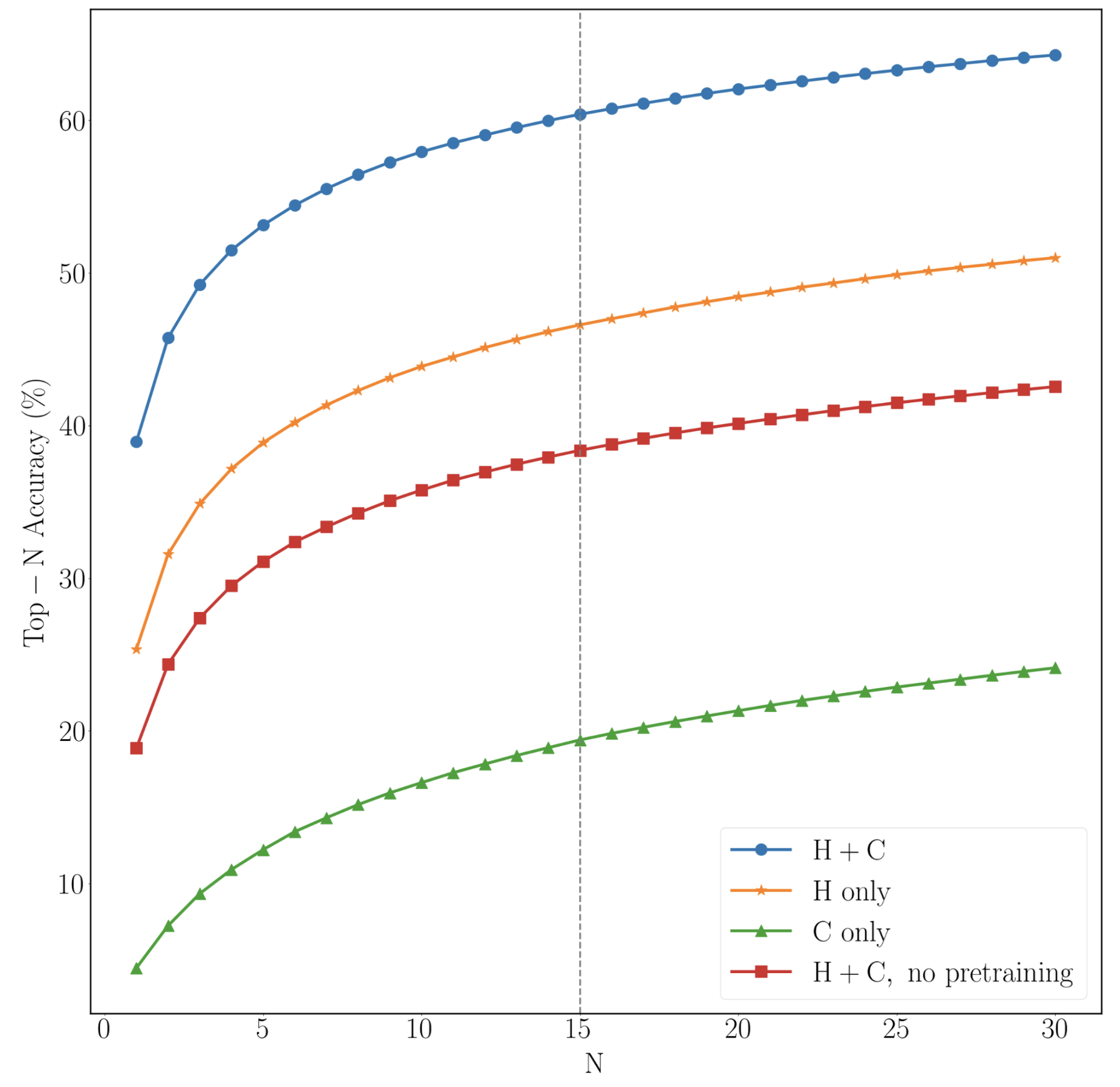}
    \caption{Top-N accuracy as a function of the number of structure predicted. The dotted line indicates the accuracy at Top-N for N=15.}
    \label{SI_fig:top_n_acc_finegrained_scan}
\end{figure}

For Top-N accuracy, we simply take the first N predictions from the model and check if the target SMILES string appears within this set of N after canonicalization. The Top-N accuracy as a function of different values of N ranging from 1 to 30 is shown in Figure~\ref{SI_fig:top_n_acc_finegrained_scan} for different combinations of inputs to the model.

\begin{figure}[!h]
    \centering
    \includegraphics[width=\textwidth]{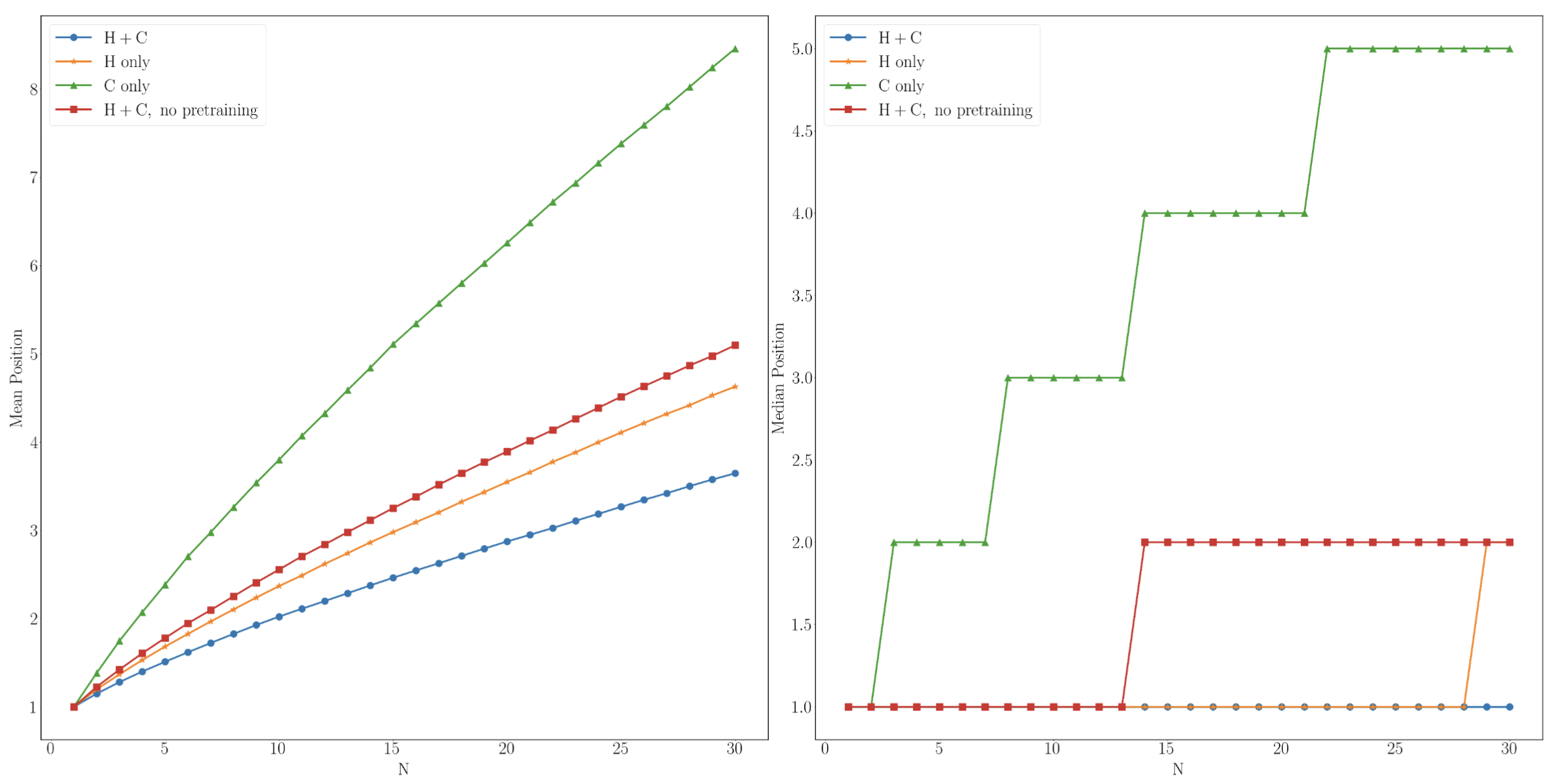}
    \caption{\newchanges{Average and median position of correct predictions among the Top-N for N from 1 to 30.}}
    \label{SI_fig:mean_med_rank_scan}
\end{figure}

\newchanges{We can also consider descriptive statistics on the position of the correct prediction for cases where the model correctly predicts the target among the Top-N predictions for different values of N. We report the maximum, minimum, mean, and median position for N = 1, 5, 10, 15, and 30 in Table~\ref{SI_table:norank_descriptive_stats} and plot the average and median positions for all N from 1 to 30 in Figure~\ref{SI_fig:mean_med_rank_scan}. We note encouragingly that for the pretrained \textsuperscript{1}H and \textsuperscript{13}C NMR combination, not only does the average position remain very low across all values of N by never rising above 5, the median position is consistently 1. Performance is systematically worse for the pretrained \textsuperscript{1}H NMR only combination and the non-pretrained \textsuperscript{1}H and \textsuperscript{13}C NMR combination across all N, but the average position still remains around 5 at N = 30 and the median position does not exceed 2 even at higher values of N. Even with the pretrained \textsuperscript{13}C NMR only combination, the median position does not exceed 5 across all N. Altogether, this indicates that if the correct molecule is predicted by the model, it is frequently predicted early on in the set of generated molecules, which is ideal since it indicates fewer required samples.}

\begin{table}[h]
\centering
\caption{\newchanges{Descriptive statistics regarding the position of the correct molecule within the Top-N for N = 1, 5, 10, 15, and 30. For every metric, lower is better.}}
\label{SI_table:norank_descriptive_stats}
\sisetup{round-mode=places}
\resizebox{\columnwidth}{!}{%
\begin{tabular}{l l l r r r r r}
\toprule
\textbf{Spectra used} & \textbf{Pretrained transformer} & \textbf{Value} ($\downarrow$) & \textbf{Top-1} & \textbf{Top-5} & \textbf{Top-10} & \textbf{Top-15} & \textbf{Top-30} \\
\midrule

\multirow[t]{4}{*}{\textsuperscript{1}H and \textsuperscript{13}C} & \multirow[t]{4}{*}{Yes}
  & Mean   & 1.00 & 1.51 & 2.02 & 2.46 & 3.65 \\
  &  & Min    &  1 & 1 & 1 &  1 & 1 \\
  &  & Max    & 1 & 5 & 10 & 15 & 30 \\
  &  & Median &  1 & 1 & 1 & 1 & 1 \\
\cmidrule{3-8}

\multirow[t]{4}{*}{\textsuperscript{1}H only} & \multirow[t]{4}{*}{Yes}
  & Mean   & 1.00 & 1.68 & 2.37 & 2.98 & 4.63 \\
  &  & Min    &  1 & 1 & 1 &  1 & 1 \\
  &  & Max    & 1 & 5 & 10 & 15 & 30 \\
  &  & Median & 1 & 1 & 1 & 1 & 2 \\
\cmidrule{3-8}

\multirow[t]{4}{*}{\textsuperscript{13}C only} & \multirow[t]{4}{*}{Yes}
  & Mean   &  1.00 & 2.39 & 3.80 & 5.11 & 8.46 \\
  &  & Min    &  1 & 1 & 1 &  1 &  1 \\
  &  & Max    & 1 & 5 & 10 & 15 & 30 \\
  &  & Median &  1 & 2 & 3 &  4 & 5 \\
\cmidrule{3-8}

\multirow[t]{4}{*}{\textsuperscript{1}H and \textsuperscript{13}C} & \multirow[t]{4}{*}{No}
  & Mean   & 1.00 & 1.78 & 2.56 & 3.25 & 5.10 \\
  &  & Min    &  1 & 1 & 1 &  1 & 1 \\
  &  & Max    & 1 & 5 & 10 & 15 & 30 \\
  &  & Median & 1 & 1 & 1 & 2 & 2 \\
\bottomrule
\end{tabular}}
\end{table}

For ranking the structure predictions from the model, we devised two methods to do so: using the binary cross entropy loss between a predicted structure's substructures and the model's substructure prediction, where lower loss is better; and using the operational log-likelihood under the model, where higher likelihood is better. 

For ranking by the binary cross entropy loss against the substructure prediction from the model, we compute for each predicted structure the following loss value:
\begin{equation}
    \mathcal{L}_{BCE}=\frac{1}{N}\sum_{i=1}^N y_i\log \hat{y}_i + (1 - y_i)\log(1-\hat{y}_i)
\end{equation}
where each ``ground truth" $y_i$ is the binary value for substructure $i$ obtained by rounding the model's predicted probability for the substructure $p_i$, and $\hat{y}_i$ is the ``predicted" binary value for substructure $i$ obtained by computing the substructure array for a predicted structure.

For ranking by the operational log-likelihood under the model, we compute for each predicted structure the following log-likelihood:
\begin{equation}
    \log p_{\theta}^k(\boldsymbol{x}|\boldsymbol{c})=\sum_{i=1}^T \log p_{\theta}^k(x_i|x_{< i}, \boldsymbol{c})
\end{equation}
where $\boldsymbol{x}$ is the predicted SMILES string with $T$ many tokens, $\boldsymbol{c}$ is the context for the model which in this case is the spectra inputs, and $p_{\theta}^k(\boldsymbol{x}|\boldsymbol{c})$ is used to denote the probability distribution parameterized by the transformer but sampled using a top-$k$ random sampling strategy\cite{fan_hierarchical_2018} with $k=5$. Note that this is not equivalent to the true log-likelihood under the model, i.e. $\log p_{\theta}(\boldsymbol{x}|\boldsymbol{c}) \neq \log p_{\theta}^k(\boldsymbol{x}|\boldsymbol{c})$, but since we consistently use the same top-$k$ sampling strategy in all of our experiments, this operational log-likelihood is sufficient for ranking the predictions. \changed{The results for both methods are tabulated below in Table~\ref{SI_table:rank_1_of_n_results}.}

\begin{table}[!ht]
    \centering
    \caption{\changed{Comparison of the performance of metrics to provide Rank-1-of-N rankings for models trained with different inputs.}}
    \label{SI_table:rank_1_of_n_results}
    \sisetup{round-mode=places}
    \large
    \resizebox{\columnwidth}{!}{\begin{tabular}{ccS[round-precision=1]S[round-precision=1]S[round-precision=1]c}
        \multicolumn{6}{c}{\textbf{\Large Ranking by substructure predictions}}\\
        \\
        \textbf{Spectra used} & \textbf{Pretrained} & \textbf{Rank-1-of-5} & \textbf{Rank-1-of-10} & \textbf{Rank-1-of-15} & \textbf{Rank-1-of-30}  \\
        \textbf{} & \textbf{transformer} & \textbf{accuracy (\%)} & \textbf{accuracy (\%)} & \textbf{accuracy (\%)} & \textbf{accuracy (\%)}\\
        \midrule
        \textsuperscript{1}H and \textsuperscript{13}C & Yes & \numericalresult{42.2} & \numericalresult{41.0} & \numericalresult{40.0} & \numericalresult{37.9}\\ 
        \textsuperscript{1}H only & Yes & \numericalresult{27.6} & \numericalresult{26.3} & \numericalresult{25.3} & \numericalresult{23.4}\\ 
        \textsuperscript{13}C only & Yes & \numericalresult{6.1} & \numericalresult{5.5} & \numericalresult{5.0} & \numericalresult{4.0}\\ 
        \textsuperscript{1}H and \textsuperscript{13}C & No & \numericalresult{23.6} & \numericalresult{23.4} & \numericalresult{23.0} & \numericalresult{21.9}\\  \midrule
        \\
        \multicolumn{6}{c}{\textbf{\Large Ranking by structure predictor probability distribution}}\\
        \\
        \textbf{Spectra used} & \textbf{Pretrained} & \textbf{Rank-1-of-5} & \textbf{Rank-1-of-10} & \textbf{Rank-1-of-15} & \textbf{Rank-1-of-30}  \\ 
        \textbf{} & \textbf{transformer} & \textbf{accuracy (\%)} & \textbf{accuracy (\%)} & \textbf{accuracy (\%)} & \textbf{accuracy (\%)}\\ 
        \midrule
        \textsuperscript{1}H and \textsuperscript{13}C & Yes & \numericalresult{47.0} & \numericalresult{48.0} & \numericalresult{48.3} & \numericalresult{48.6}\\ 
        \textsuperscript{1}H only & Yes & \numericalresult{33.2} & \numericalresult{34.3} & \numericalresult{34.6} & \numericalresult{34.9}\\ 
        \textsuperscript{13}C only & Yes & \numericalresult{10.4} & \numericalresult{12.5} & \numericalresult{13.5} & \numericalresult{14.5}\\ 
        \textsuperscript{1}H and \textsuperscript{13}C & No & \numericalresult{26.6} & \numericalresult{28.1} & \numericalresult{28.8} & \numericalresult{29.6}\\  \bottomrule
    \end{tabular}}
\end{table}

\subsection{Fine-tuning on experimental spectra}\label{SI_subsec:experimental_spectra_runs}

\subsubsection{Data curation}
\begin{figure}[!h]
    \centering
    \includegraphics[width=\textwidth]{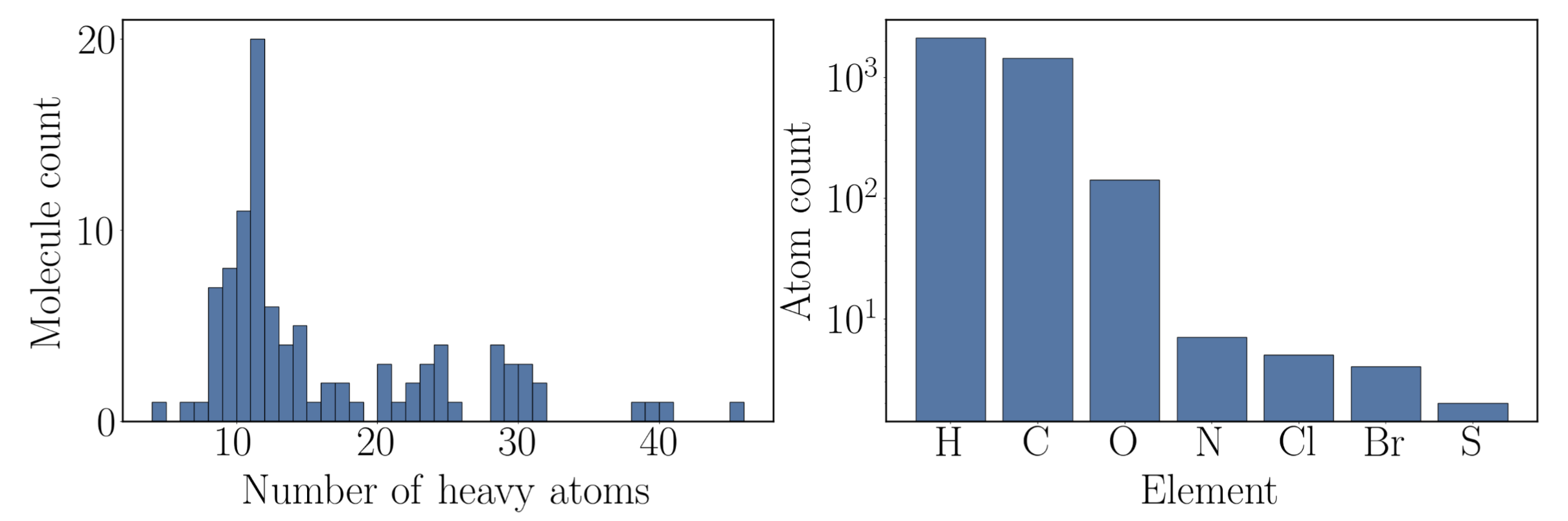}
    \caption{Size and element distribution for the 100 experimental spectra used.}
    \label{SI_fig:exp_spectra_dists}
\end{figure}

For adapting our framework to experimental spectra, we obtained a set of 100 spectra from the Biological Magnetic Resonance Data Bank (BMRB). We filtered the BMRB spectra specifically for those collected at \changed{field strengths} of 400 to 500 MHz and with CDCl$_3$ as the solvent, and looked for entries that contained both \textsuperscript{1}H and \textsuperscript{13}C NMR. After removing repeated molecules based on the SMILES strings obtained from the InChi codes, we were left with a total of 100 spectra, all collected on Bruker spectrometers. The element and size distributions of the molecules are shown in Figure~\ref{SI_fig:exp_spectra_dists}.

For the \textsuperscript{1}H NMR, we applied the following minimal pre-processing steps:
\begin{enumerate}
    \item Apply a Bernstein polynomial baseline correction using MestReNova\cite{willcott_mestre_2009} with the default parameters.
    \item Interpolate the spectrum onto our standard shift grid of 28000 values which evenly spans from -2 to 12 ppm.
    \item Normalize the intensities such that each value is between 0 and 1.
    \item Remove any intensity values lower than 0.001 by setting them to zero, thus removing most of the remaining baseline noise.
    \item Zero the spectrum between -0.1 ppm and +0.1 ppm to remove reference peaks.
    \item Re-normalize the spectrum to intensities between 0 and 1.
\end{enumerate}

This procedure does not include the manual removal of any impurity or solvent peaks. For the \textsuperscript{13}C NMR spectra, we \changed{format the chemical shifts into a vector directly}, but removed any reference chemical shifts that had a value less than 0.25 ppm.

\subsubsection{Supervised fine-tuning}

In order to fine-tune our model on this collection of 100 experimental spectra, we used a supervised fine-tuning strategy that optimized only a subset of the model's parameters. We first partitioned the 100 spectra into 50 training examples, 25 validation examples, and 25 testing examples using a scaffold splitting based on the molecules' Murcko Scaffolds\cite{bemis_properties_1996} extracted using RDKit. We then trained our multitask model on the small fine-tuning training set starting from our \textsuperscript{1}H + \textsuperscript{13}C model trained on the simulated spectra. During this training, all of the parameters were frozen except for a certain number of layers in the transformer decoder of the structure elucidation branch of the multitask model. To select the optimal number of layers to be unfrozen, we unfroze a variable number of decoder layers starting with the later layers and then moving towards the earlier layers. The intuition behind this approach is that the convolutional embedding and transformer encoder have already learned meaningful and transferable features extracted from the various simulated NMR spectra used during the large scale training, and we are simply trying to adapt the model to a different domain by tuning the weights in the later layers of the model which are more consequential for downstream adaptation tasks. 

We trained different variations of the model with different numbers of decoder layers unfrozen, ranging from 1 layer to a maximum of 4 layers, and our results are summarized in Table~\ref{SI_table:SFT_ablations}. All models were trained using the AdamW optimizer with a lower learning rate of $1.0\times 10^{-6}$ than used in training the models from scratch on the simulated data. The substructure weighting factor was set to 0, i.e. the loss is entirely determined by the structure prediction. We trained each of the model variations here until convergence by monitoring overfitting on the fine-tuning validation set. We used early stopping to select the best model which we then evaluated on the fine-tuning test set. Due to the small number of spectra involved, we evaluated the structure accuracy of the model variations 30 different times, with a different random seed for each, and report the average structure accuracy alongside the standard error of the mean of $\sigma/\sqrt{n}$. The structure accuracy is computed using 15 generated predictions from the model. 

\begin{table}[!h]
    \centering
    \caption{Structure accuracies of models with different numbers of decoder layers unfrozen on the experimental test set. The standard error in the mean (computed from 30 different evaluations with a different random seed each time) is reported alongside the average.}
    \label{SI_table:SFT_ablations}
    \sisetup{round-mode=places}
    \large
    \resizebox{\columnwidth}{!}{\begin{tabular}{cS[round-precision=2]S[round-precision=2]}
        \toprule
        \textbf{Number of layers} & \textbf{Number of trainable parameters} & \textbf{Top-15 structure accuracy (\%)} \\ \midrule
        1 & 12.8 M & 8.40  $\pm$ 0.78 \\ 
        2 & 25.4 M & 14.80 $\pm$ 0.79 \\ 
        3 & 38.0 M & 20.13 $\pm$ 1.01 \\ 
        4 & 50.6 M & 21.47 $\pm$ 1.06 \\ \midrule
    \end{tabular}}
\end{table}

\begin{figure}[!h]
    \centering
    \includegraphics[width=\textwidth]{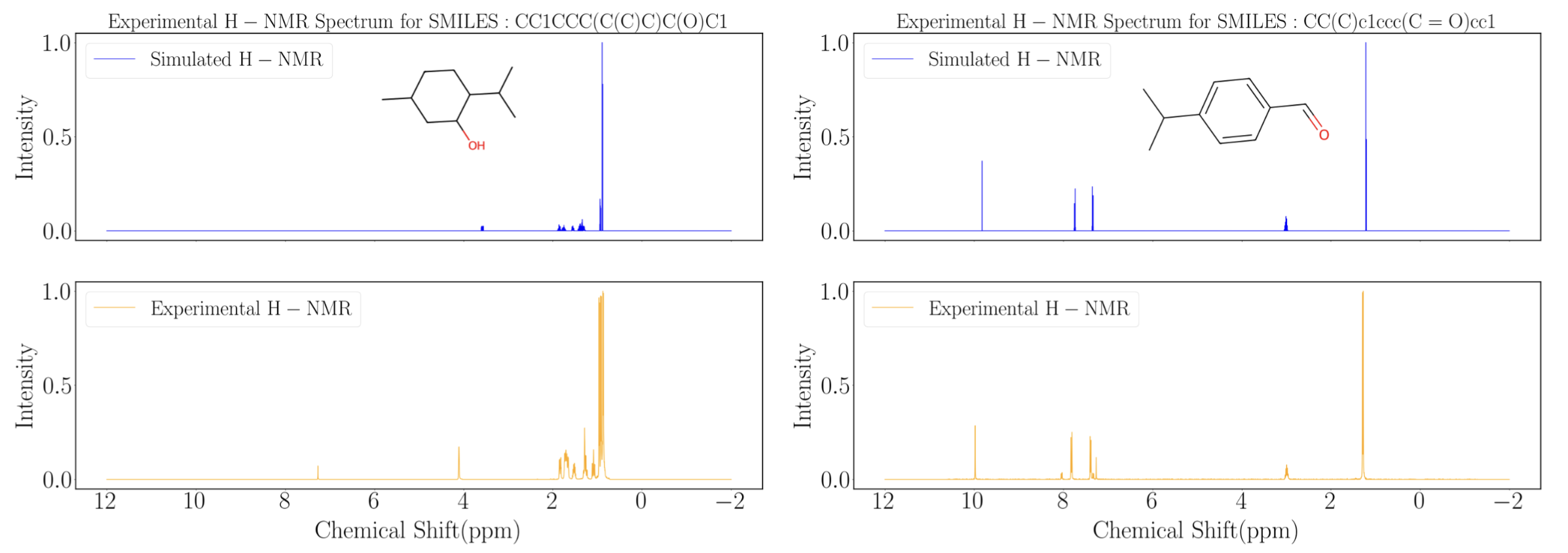}
    \caption{Comparison of simulated and experimental \textsuperscript{1}H NMR spectra for two correctly predicted examples, where the simulated spectra is generated using the ACD/Labs predictor and the experimental spectra is obtained from the BMRB and preprocessed according to our workflow.}
    \label{SI_fig:comparison_sim_exp_correct_preds}
\end{figure}

Using the model with the highest structure accuracy on the experimental spectra, which for this data was the model with \changed{4 layers unfrozen (Table~\ref{SI_table:SFT_ablations} row 4)}, we also evaluated the model's performance on the original test set of simulated NMR spectra to assess how much the model's performance on the simulated spectra degraded as a result of the fine-tuning procedure. The model's performance does not show a significant decrease, maintaining \changed{a Top-15 accuracy of \numericalresult{60.2\%}} on the original test set of simulated spectra, indicating that the fine-tuning protocol helped the model adapt to the out of domain experimental data without inducing catastrophic forgetting. This indicates that our multitask framework shows promise as a potential foundational model, enabling inexpensive and non-destructive fine-tuning to a variety of downstream tasks and domains. Examples of some of the molecules the model predicted correctly are provided in Figure~\ref{SI_fig:comparison_sim_exp_correct_preds} alongside their simulated \textsuperscript{1}H NMR spectra generated using the v2024.2 batch NMR predictors from ACD/Labs and the experimental \textsuperscript{1}H NMR spectra after the processing steps outlined above. This shows that the model is learning to adapt to the additional features present in the experimental spectra.

\subsection{Additional figures and analysis}

\begin{figure}[!h]
    \centering
    \includegraphics[width=\textwidth]{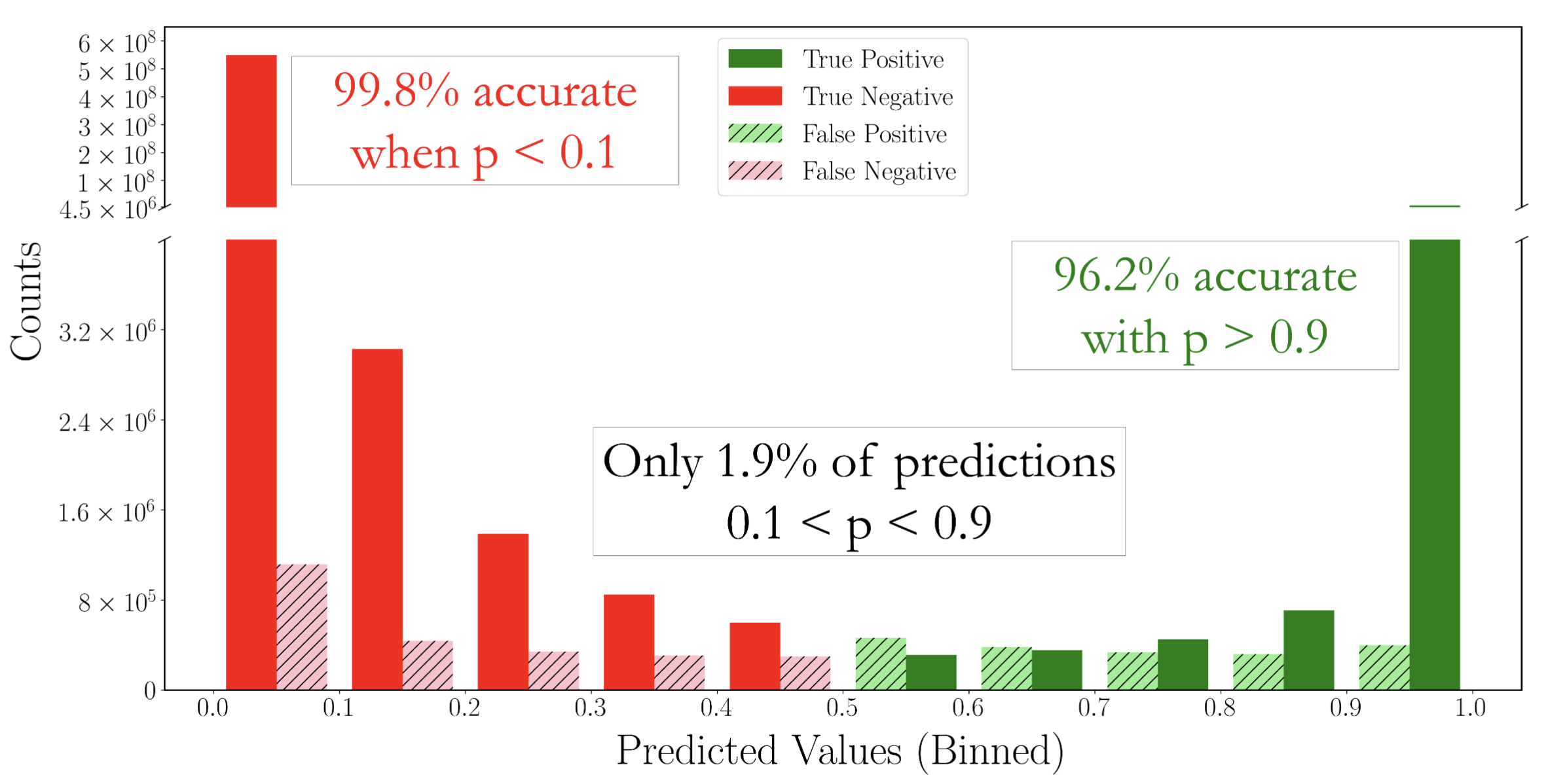}
    \caption{Multitask model substructure prediction performance resolved by predicted probability and true/false positive/negatives.}
    \label{SI_fig:spec_to_struct_substruct_acc_breakdown}
\end{figure}

\begin{figure}[!h]
    \centering
    \includegraphics[width=\textwidth]{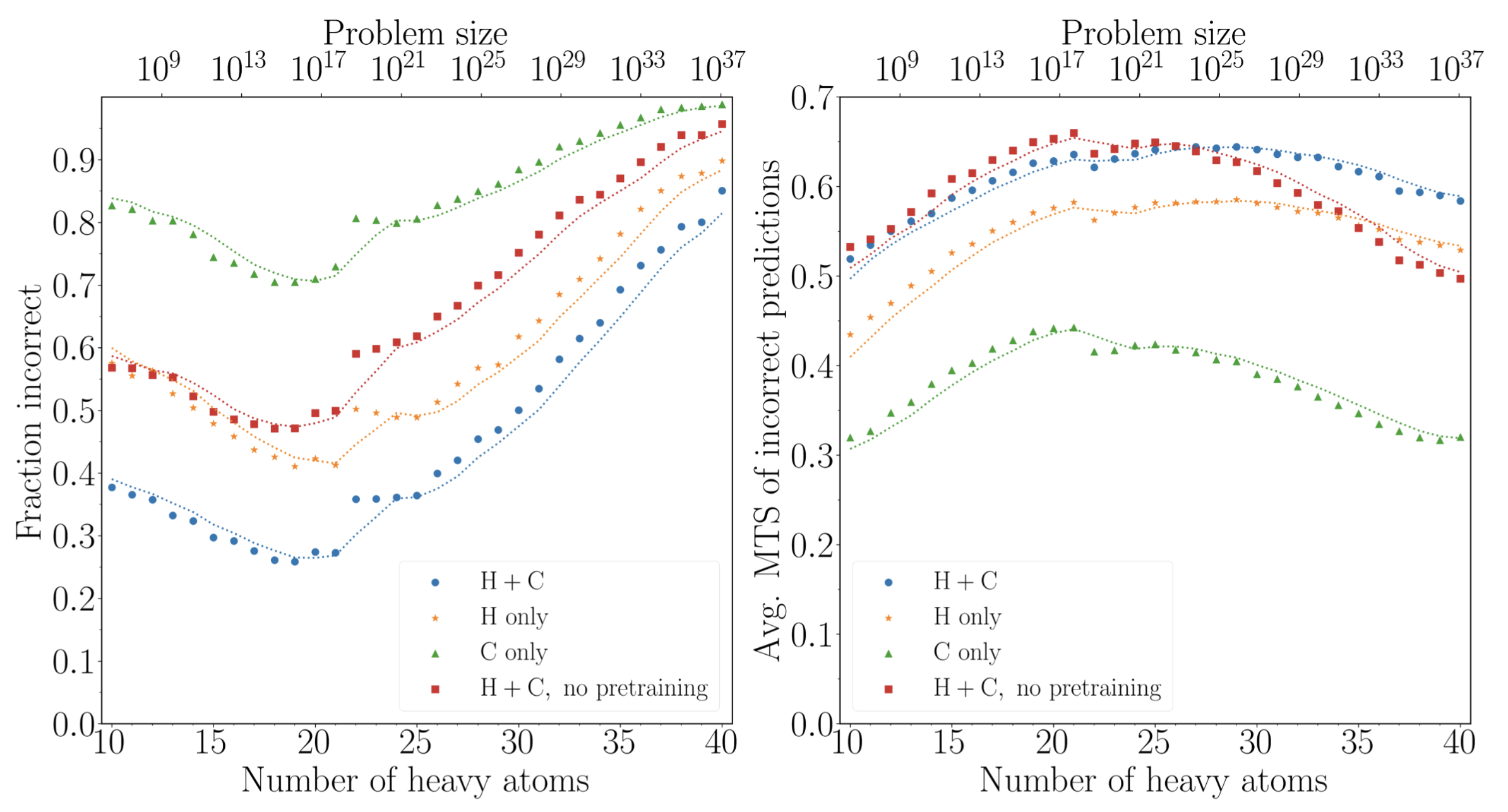}
    \caption{Analysis of (left) structure accuracy as a function of the molecule size and the spectral modalities used and (right) average MTS value as a function of molecule size and spectral modalities used. The dashed line in both figures denote the running average.}
    \label{SI_fig:spec_to_struct_ablation_size_breakdown}
\end{figure}

\begin{figure}[!h]
    \centering
    \includegraphics[width=\textwidth]{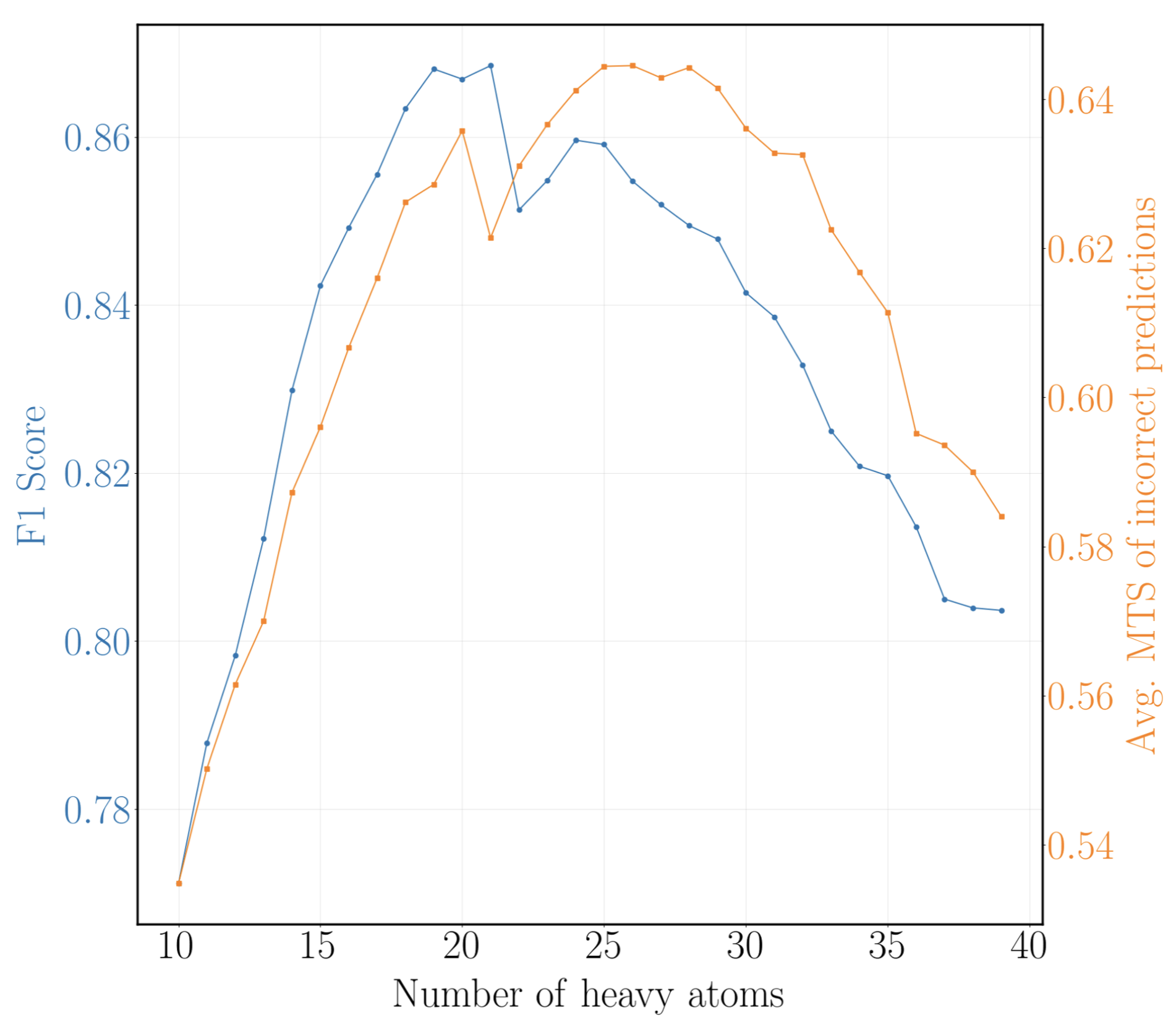}
    \caption{The F1 score for substructure prediction from the multitask framework for different molecule sizes and the average MTS of incorrect predictions. The two quantities are positively correlated, with $R^2 = 0.73$.}
    \label{SI_fig:spec_to_struct_F1_and_MTS_corr}
\end{figure}

\begin{figure}[!h]
    \centering
    \includegraphics[width=\textwidth]{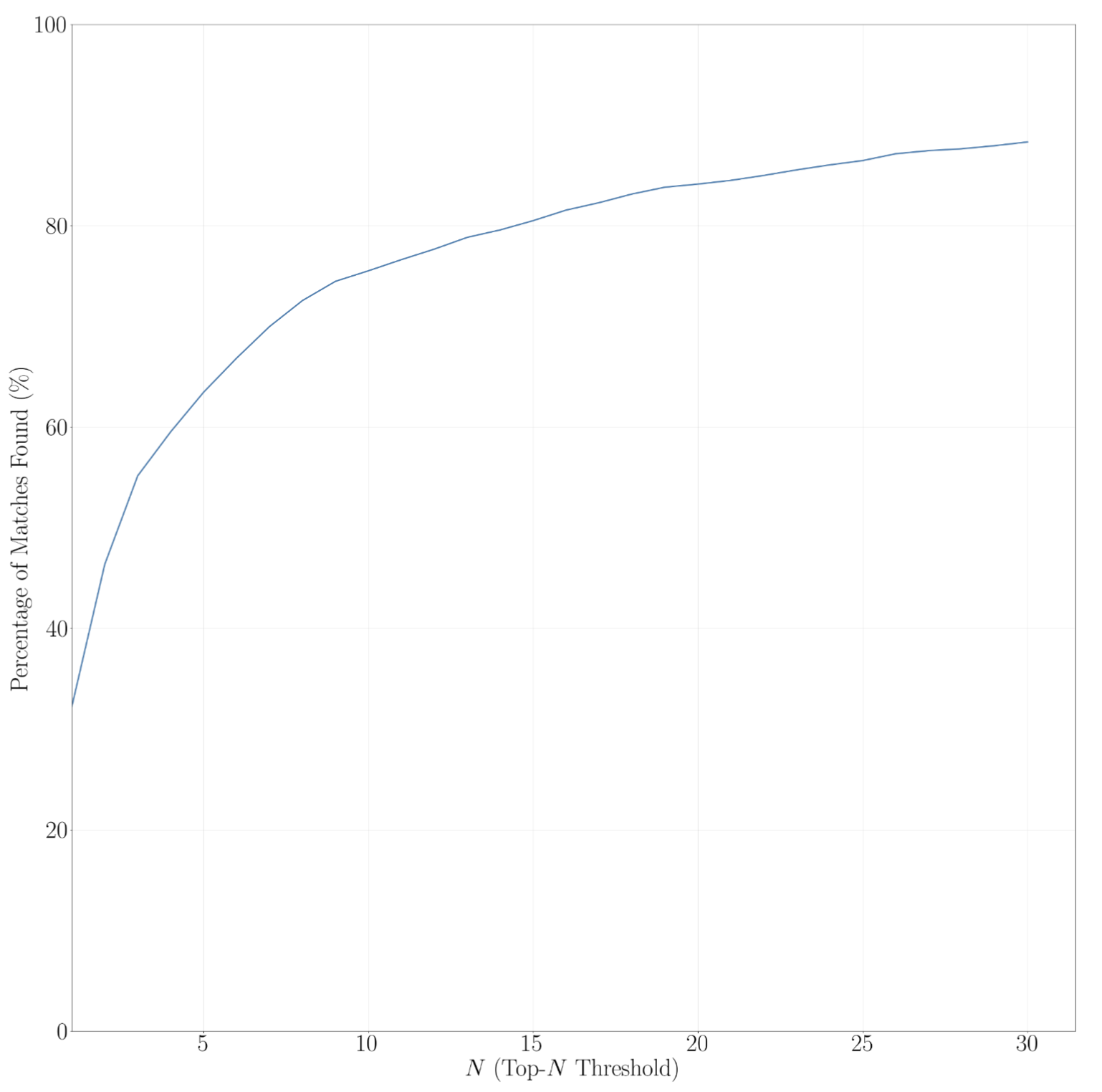}
    \caption{\newchanges{The percentage of 40 heavy atom molecules captured at different Top-N thresholds for N from 1 to 30.}}
    \label{SI_fig:dataset_search_threshold}
\end{figure}

\clearpage
\bibliography{references}